\documentclass[superscriptaddress,reprint,amssymb, nobibnotes, aps, pra]{revtex4-2}
\usepackage{graphicx}
\usepackage{amsmath}
\usepackage{float}
\usepackage{braket}
\usepackage{bibentry}

\usepackage{color,soul}
\usepackage{appendix}

\begin{document}

\title{Encoding quantum optical phase-space with classical wireless microwave constellation}

\author{Niloy Ghosh}
\author{Sarang Pendharker}

\affiliation{Department of Electronics $\&$ Electrical Communication Engineering, Indian Institute of Technology Kharagpur, W. Bengal, 721302, India.}

\begin{abstract}
This paper develops a theoretical framework for enabling seamless transfer of digital information from classical microwave domain to the quantum optical domain in wireless-to-optical converters. A quantum mechanical network model is introduced to characterize microwave-to-optical digital information mapping in antenna-coupled electro-optic modulator-based converters. Design guidelines are discussed to maximize the information mapping strength. The derived model is then extended to show phase-space encoding of optical coherent-states with classical wireless microwave constellation. Further, the challenge of inter-symbol overlap in the encoded quantum optical phase-space due to quadrature fluctuations is highlighted. The possibility of erroneous phase-space encoding due to quadrature fluctuations is pointed out, followed by a potential mitigation technique. The presented framework also lays the groundwork for encoding other non-classical states of light such as squeezed states, and hence forms the basis for bridging classical microwave and quantum optical links in the near future.
\end{abstract}

\maketitle

\section{Introduction}

Seamless bridging of wireless and optical technologies can potentially enable the implementation of high-bandwidth communication networks with enhanced coverage area \cite{zhu2024high,filgueiras2023wireless,lopez2021power,lim2019evolution,jia20180}. Such networks are typically composed of short-range wireless microwave cells interfaced with long-range optical-fiber links \cite{n2023radio,akyildiz2022terahertz,burla2019500, harter2019wireless, nagatsuma2016advances}. The performance of such networks significantly relies on the efficiency of seamless Wireless-to-Optical (W-O) converters functioning at the interface between wireless microwave and optical domains \cite{ummethala2019thz}. Typically, seamless W-O converters are antenna-coupled electro-optic modulators that modulate optical signals with microwave signals. Such converters are highly desirable for next-generation communication systems owing to their low end-to-end latency and simplistic design, as compared to conventional electronics-based W-O converters \cite{li2014fiber,zhang2022real,cai2023real,zhang2022real}. 

Despite appreciable progress in the development of seamless W-O converters over the years, their integration with digital communication links remains vastly unaddressed \cite{horst2022transparent, wijayanto2013electrooptic, kaji2021w, murata2021antenna, park2015free,miyazeki2020ingaas}. To understand this, it must first be reiterated that most modern wireless communication systems predominantly transmit information by encapsulating digital information as symbols in the amplitude or phase of microwave wireless electric-fields \cite{haykin1988digital}. Therefore, in order to integrate seamless W-O converters with digital links, it is important to first investigate how the digital symbols encapsulated within  microwave electric fields can be translated to the optical domain. However, this aspect remained unexplored until we proposed a novel framework called bi-layered electro-optic modulation for seamless microwave-to-optical digital symbol-mapping for the first time \cite{Ghosh_2021, 10107736}.  Recently, we extended our proposed symbol mapping model to offer architectural solutions for enabling photonic detection of digital symbols encoded in wireless microwave constellations for the first time \cite{10535154}. 

At this juncture, it must be highlighted that so far only classical nature of light has been considered for modeling seamless microwave-to-optical digital symbol mapping. However, this consideration may not be generally sufficient to model microwave-to-optical digital symbol mapping when quantum effects of light come into play. Therefore, such cases necessitate effects such as superposition, interference, quantum noise, etc, to be taken into consideration. Therefore, establishing a theoretical framework for seamless transfer of digital information from classical microwave to the quantum optical domain is imperative. 

The development of such a theoretical framework necessitates deriving a quantum mechanical model for electro-optic modulation as a prerequisite. Numerous studies have been performed on the analog modulation of quantum optical states with continuous-wave classical microwave E-fields. A thorough description of electro-optic phase-modulation considering the modulated optical light to be quantum was presented \cite{capmany2010quantum}. An exactly solvable multi-mode quantum model for electro-optic phase-modulation was reported using an algebraic framework \cite{miroshnichenko2017algebraic}. Furthermore, a quantum mechanical framework for electro-optic phase-modulation considering the perturbative effects of dispersion was outlined \cite{horoshko2018quantum}. The impact of classical microwave phase-noise in a modulated quantum optical mode was also studied \cite{pratap2020quantum}. Recent research also reported the quantum mechanical description of a dual-drive Mach-Zehnder modulator for coherent quantum optical communication links \cite{pratap2023photon}. Furthermore, the experimental characterization of quantum optical sideband generation in electro-optic phase-modulation was carried out through a two-photon interference experiment \cite{imany2018characterization}. However, despite the appreciable progress, the scenario in which the modulating classical microwave E-field is digitally modulated has been vastly overlooked, for the existing models solely consider the received microwave E-field to be unmodulated i.e. a pure sinusoid. 

This paper aims to lay the theoretical framework behind classical microwave to quantum optical digital symbol mapping in seamless wireless-to-optical converters for the first time. This paper is organized as follows. In section~II, a quantum mechanical network model for characterizing seamless electro-optic phase-modulation of a quantum state of light with digitally modulated classical microwave E-field is derived. In section~III, the derived framework is extended to model the transfer of digital symbols from classical microwave constellation to quantum optical phase-space of optical coherent-states. This lays the foundation for phase-space encoding of quantum optical light with classical microwave constellation. An equivalent unitary quantum operator is derived to show that phase-shift-keying in the classical microwave domain causes a rotation operation in the encoded quantum optical phase-space. Furthermore, the challenge of intersymbol overlap in the encoded quantum optical phase-space due to quadrature fluctuations is highlighted. The possibility of erroneous phase-space encoding due to quadrature fluctuations is pointed out, followed by a potential mitigation technique. The framework established in this paper opens the avenue for encoding and manipulating other quantum optical states such as squeezed-states, and will have a crucial role in bringing classical and quantum communication links in the future.

\section{Digitally modulated classical microwave to quantum optical conversion}

\subsection{Electro-optic interaction Hamiltonian}
Figure~1 illustrates a seamless W-O converter composed of an antenna-coupled electro-optic phase-modulator. The converter comprises a Lithium Niobate (LiNbO$_3$) optical waveguide channelized through a centrally slotted patch antenna that functions as the modulating element. The waveguide dimension was chosen as 608nm x 164nm to enable single-mode optical wave propagation of wavelength 1555$nm$. The effective refractive index $n_{op}$ of the optical waveguide was computed to be 1.734 using effective-index method \cite{10107736}. We considered an $x$-polarized quantum optical mode of frequency $\omega_{op}$ propagating in the $z$-direction to be launched at the input end of the waveguide. The electric-field operator $\hat{E}_{op}(t)$ associated with the quantum optical mode can be expressed as \cite{gerry2023introductory},
\begin{equation}
    \hat{E}_{op}(t)  = j \sqrt{\frac{\hbar \omega_{op}}{2\epsilon_o \epsilon_{op} V}}\ \bigg( \hat{a}\ e^{-j(\omega_{op}t-k_{op}z) } - \hat{a}^{\dagger}\ e^{j(\omega_{op}t-k_{op}z) } \bigg)
\label{eq1}
\end{equation}
where $k_{op}$ is the phase-constant of the quantum optical mode,  $\epsilon_o$ is the permittivity of vacuum, and $\epsilon_{op}$ is the effective relative permittivity of the optical waveguide, and $V$ is the effective modal volume of the quantum optical mode in the waveguide. Also, $\hat{a}^{\dagger}$ and $\hat{a}$ are the creation and annihilation operators, respectively. In the absence of classical wireless microwave reception, the total Hamiltonian of the quantum optical mode traveling in the converter is equal to the free Hamiltonian $\hat{H}_o$ of a quantum harmonic oscillator \cite{bransden2000quantum},
\begin{equation}
        \hat{H}_o= \hbar \omega_{op} \hat{a}^{\dagger}\hat{a}
\label{eq2}
\end{equation}
\begin{figure}[ht]
    \centering
\includegraphics[width=\linewidth]{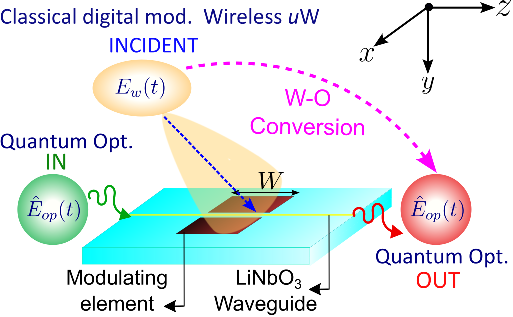}
    \caption{Digitally modulated classical wireless microwave to quantum optical conversion.}
    \label{Fig_2}
\end{figure}
We next consider an $x$-polarized classical wireless microwave E-field $E_w(t)$ to be received by the converter,
\begin{equation}
    E_w(t) = |E_w|\ \sin{(\omega_{w}t + b_i)}
\label{eq3}
\end{equation}
where $|E_w|$ and $\omega_w$ is the field-strength and frequency of $E_w(t)$, respectively. We consider $E_w(t)$ to be digital phase-shift-keying (PSK) modulated \cite{haykin1988digital}, where $b_i$ is the phase associated with the $i^{th}$ digital symbol encapsulated in $E_w(t)$. On receiving $E_w(t)$, an $x$-polarized E-field of enhanced field-strength gets induced in the centrally slotted modulating element present in the converter\cite{Ghosh_2021}. Consequently, Pockel's electro-optic effect perturbs $\epsilon_{op}$ of the LiNbO$_3$ optical waveguide section channelized through the modulating element. Considering the optic-axis of the LiNbO$_{3}$ waveguide to be oriented along the $x$-axis, the modified relative permittivity of the waveguide section due to Pockel's effect can be expressed as,
\begin{equation}
    \epsilon'_{op}(t) = \epsilon_{op} + \delta \epsilon_{op}(t)   = \epsilon_{op} - \epsilon_{op}^{2} r_{33} \gamma E_w(t)
\label{eq4}
\end{equation}
where $\delta \epsilon_{op}(t)$=$-\epsilon_{op}^{2} r_{33} \gamma E_w(t)$ is the perturbed relative permittivity of the waveguide section, $r_{33}$ is the Pockel's electro-optic coefficient of LiNbO$_3$, and $\gamma$ is the slot field-strength enhancement factor of the modulating element. The shift in relative permittivity $\delta \epsilon_{op}(t)$ of the waveguide section perturbs the Hamiltonian of the quantum optical mode by $ \hat{H}_I(t)$ {(see Appendix-A)},
\begin{equation}
        \hat{H}_I(t) = - \frac{\epsilon_{op} r_{33}}{2} \gamma E_w(t)\ \hbar \omega_{op} \hat{a}^{\dagger}\hat{a}
\label{eq6}
\end{equation}
It can be observed that the shift in Hamiltonian originates from the interaction between the quantum optical mode and the received classical microwave electric-field mediated by Pockel's effect in the waveguide section. The time-dependence of $\hat{H}_I(t)$ is governed by the  classical microwave electric-field $E_w(t)$. Therefore, the net Hamiltonian $\hat{H}(t)$ of the quantum optical mode becomes time-dependent and can be expressed as,
\begin{equation}
        \hat{H}(t) =  \hat{H}_o + \hat{H}_I(t) = \hbar \omega_{op} \hat{a}^{\dagger}\hat{a}- \frac{\epsilon_{op} r_{33}}{2} \gamma E_w(t)\ \hbar \omega_{op} \hat{a}^{\dagger}\hat{a}
\label{eq6.1}
\end{equation}
Therefore, the Hamiltonian quantum optical mode can be modulated by the received classical microwave E-field in the W-O converter. In the upcoming subsection, we extend the discussion to derive the time evolution of the modulated quantum optical state from Schrodinger's picture.

\subsection{Evolution of modulated quantum optical state}

The time evolution of the modulated quantum optical state $\ket{\psi}$ in the W-O converter is governed by time-dependent Schrodinger's equation as follows,
\begin{equation}
j\hbar \frac{d\ket{\psi(t)}}{dt} = \hat{H}(t) \ket{\psi(t)}
\label{eq8}
\end{equation}

The general solution of time-dependent Schrodinger's equation is of the following form,
\begin{equation}
\ket{\psi(t)} = \sum_{k=0}^{\infty} \ket{\psi_k(t)} = \sum_{k=0}^{\infty} C_k(t) \ket{k} 
\label{eq9}
\end{equation}
where $C_k(t)$ is the time-evolving probability amplitude of the $k^{th}$ Fock basis-state $\ket{k}$ associated with the quantum optical state $\ket{\psi(t)}$. By substituting of $\ket{\psi(t)}$ from Eq.~(\ref{eq9}) into  Eq.~(\ref{eq8}), we get the following differential equation,
\begin{equation}
    j \hbar \sum_{k=0}^{\infty} \frac{d C_k(t)}{dt} \ket{k}  = \sum_{k=0}^{\infty} \hat{H}(t) C_k(t)\ket{k}
\label{eq10}
\end{equation}
We then substitute the expression of $\hat{H}(t)$ from Eq.~(\ref{eq6.1}) to the above equation, and derive the differential equation governing the evolution of the probability amplitude $C_k(t)$ of the $k^{th}$ Fock basis-state as,
\begin{equation}
    j \hbar \frac{d C_k(t)}{dt} = k \hbar \omega_{op} \bigg(1-  \frac{\epsilon_{op} r_{33}}{2}\ \gamma E_w(t)\bigg) C_k(t)
\label{eq10_2_1}
\end{equation}
Likewise, the evolution of the probability amplitude $C_1(t)$ of the single photon Fock basis-state $k$=1 is governed by the following differential equation,
\begin{equation}
    j \hbar \frac{d C_1(t)}{dt} = \hbar \omega_{op} \bigg(1-  \frac{\epsilon_{op} r_{33}}{2}\ \gamma E_w(t)\bigg) C_1(t)
\label{eq10_2}
\end{equation}
\begin{figure}[h]
    \centering
\includegraphics[width=\linewidth]{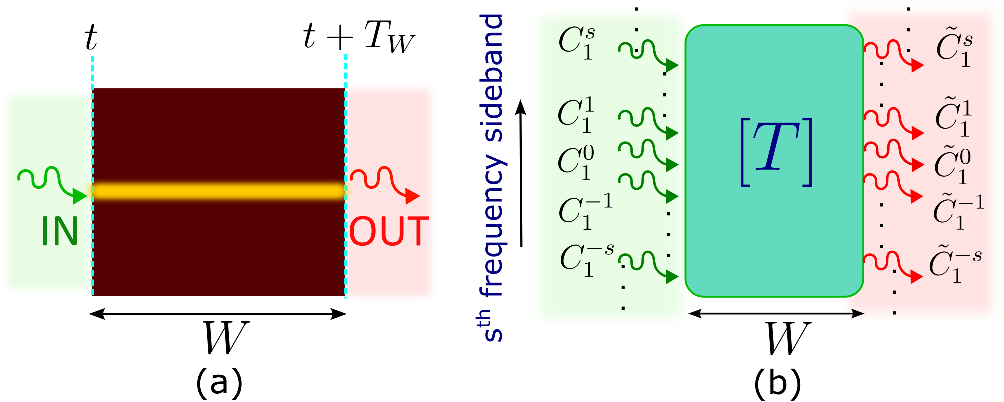}
    \caption{(a) Modulating element of width $W$ in the W-O converter. (b) Equivalent quantum mechanical network model of the modulating element.}
    \label{Fig2}
\end{figure}

The time-evolving probability amplitude $C_1(t)$ can be derived by integrating Eq.~(\ref{eq10_2}) in the time window during which the quantum optical state is modulated by the received classical microwave E-field in the converter. We consider the quantum optical state to enter the modulating element of width $W$ at the time instant $t$, and exit at the time instant $(t+T_W)$, as depicted in Fig.~\ref{Fig2}(a). Here $T_W$=$(W \sqrt{\epsilon_{op}}/c)$ is the time taken by the quantum optical state to travel a distance of $W$ in the waveguide. The probability amplitude $C_1(t)$ at the output of the modulating element can be derived to be {(see Appendix-B)},
\begin{equation}
\Tilde{C}_{1}= C_{1}\ e^{-j(\omega_{op}t+k_{op} W + \theta_i(t))}
\label{eq12}
\end{equation}
where $\Tilde{C}_{1}$ is the probability amplitude of the modulated quantum optical state, and $C_1$ is its probability amplitude of the input quantum optical state. Also, $\theta_i(t)$ is the phase encoded in the complex probability amplitude of the modulated quantum optical state, and is related to the digital phase $b_i$ of the received classical M-PSK microwave E-field as follows,
\begin{equation}
\theta_i(t) = \delta \theta \sin{(\omega_w t + \phi + b_i )}
\label{eq12_2}
\end{equation}
where $\delta \theta$ represents the electro-optic modulation-depth, which can be expressed as follows,
\begin{equation}
\delta \theta = -\frac{\omega_{op}\epsilon_{op} r_{33} \gamma}{\omega_w}\  \sin{\bigg(\frac{\omega_w \sqrt{\epsilon_{op}} W}{2c} \bigg)}\ |E_{w}| 
\label{eq12_3}
\end{equation}
Also the offset phase term $\phi$ in Eq.~(\ref{eq12_2}) can be expressed as,
\begin{equation}
\phi = \frac{\omega_w \sqrt{\epsilon_{op}} W}{2c}
\label{eq12_4}
\end{equation}
Substituting the expression of $\theta_i(t)$ from Eq.~(\ref{eq12_2}) in Eq.~(\ref{eq12}), and expanding the resultant equation 
using Jacobi-Anger expansion, we get the following expression,
\begin{equation}
\Tilde{C_{1}} = C_{1} e^{-j(\omega_{op}t + k_{op}W)} \sum^{s=-\infty}_{s=\infty} j^{s} J_s(\delta \theta)\ e^{-j s({\omega_w}t +  \phi + b_i)}
\label{eq15}
\end{equation}
\begin{figure*}[ht]
    \centering
\includegraphics[width=\linewidth]{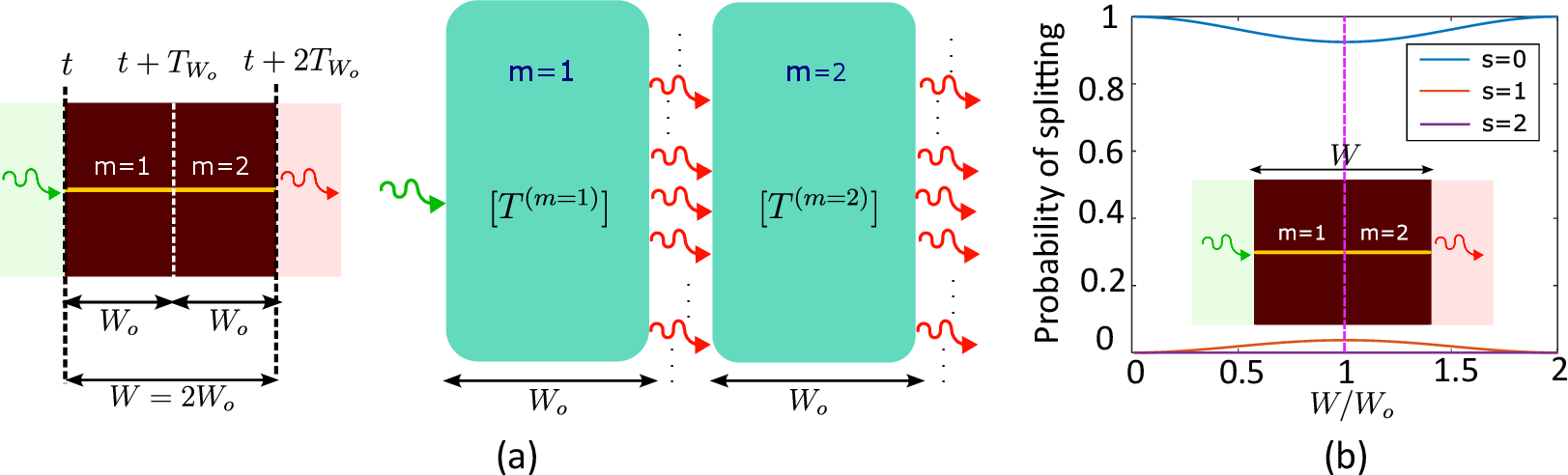}
    \caption{(a) Equivalent quantum mechanical network model of a modulating element of width $W$=$2W_o$. (b) Variation of probability amplitude splitting of a modulated optical photon with varying modulating element width, considering a classical microwave reception of field-strength $|E_w|$=50V/m.}
    \label{Fig4}
\end{figure*}
Equation~(\ref{eq15}) indicates that the probability amplitude of the quantum optical state initially evolving at frequency $\omega_{op}$ splits into various sidebands evolving at frequencies $(\omega_{op} \pm s\omega_w)$ (where $s$=0,1,2,3,...) upon being modulated by the received classical microwave E-field. Therefore, the modulating element can be ideally modeled as an infinite-port quantum mechanical network, as depicted in Fig~\ref{Fig2}(b). In the network, each port denoted by ${C^{s}_{1}}$ represents the $s^{th}$ sideband probability amplitude of the quantum optical state ( single photon Fock basis-state) at the input of the network. Similarly, each port denoted by $\Tilde{C^{s}_{1}}$ represents the $s^{th}$ sideband probability amplitude of the modulated quantum optical state at the output of the network. The output probability amplitude vector $[\Tilde{C^{s}_{1}}]$ can be related to the input probability amplitude vector $[{C^{s}_{1}}]$ by an equivalent infinite-dimensional transmission matrix $[T]$ {(see Appendix-B)}.
%\begin{equation}
    %[T] = e^{-jk_{op}W} \begin{bmatrix}
          %\textbf{.} & \textbf{.} & \textbf{.} & \textbf{.} & \textbf{.} \\
        %\textbf{.}& J_0(\delta \theta) & j^{-1} J_{-1} (\delta \theta) e^{j\gamma} & j^{-2} J_{-2}(\delta \theta)  e^{j2\gamma} & \textbf{.} \\
         %\textbf{.}&  j^{} J_1(\delta \theta) e^{-j\gamma} & J_0(\delta \theta) & j^{-1} J_{-1}(\delta \theta)  e^{j\gamma} & \textbf{.} \\
         %\textbf{.} & j^{2} J_2(\delta \theta) e^{-j2\gamma} & j^{1} J_1(\delta \theta) e^{-j\gamma} & J_0(\delta \theta) & \textbf{.} \\
         %\textbf{.} & \textbf{.} & \textbf{.} & \textbf{.} & \textbf{.} \\
        %\end{bmatrix} 
%\label{eq18_2}
%\end{equation}
 Therefore, at the output of the modulating element, the optical photon can remain at the center frequency $\omega_{op}$ with probability $J^{2}_0(\delta \theta)$, or shift to sideband frequency $(\omega_{op} \pm s\omega_w)$ with probability $J^{2}_{\pm s}(\delta \theta)$. The extent of probability amplitude splitting is an indicator of the electro-optic modulation-depth, which in turn is the measure of the modulation-strength.

In our earlier works, we had shown that the modulation-depth can be maximized by suitably choosing the modulating-element width and element periodicity, considering the modulated light to be classical \cite{Ghosh_2021, 10107736,10535154}. In the upcoming subsections, we show this for the quantum case, using the derived network model.

\subsection{Maximizing modulation-strength in a single modulating element}

From Eq.~(\ref{eq12_3}), it can be found that the electro-optic modulation-depth $\delta \theta$ can be maximized if the modulating element width is chosen as $W$=$W_o$=$(\pi c/2 \omega_w \sqrt{\epsilon_{op}})$. The maximized electro-optic modulation-depth $\delta \theta$=$\delta \theta_o$ in that case becomes equal to, 
\begin{equation}
    \delta\theta_o = -\frac{\omega_{op}\epsilon_{op} r_{33} \gamma}{\omega_w}\ |E_{w}|
\label{eq27}
\end{equation}
Additionally, it can also be found that if the width is chosen as twice the optimum width i.e. $W$=$2W_o$=$(\pi c/\omega_w \sqrt{\epsilon_{op}})$, the electro-optic modulation-depth $\delta \theta$ becomes zero. This can be physically explained using the previously derived quantum mechanical network model. To do this, we first consider the physical width $W$ of the modulating element to be $W$=$2W_o$. We then divide the entire modulating element into two individual sections of width $W_o$ cascaded to each other, as shown in Fig.~\ref{Fig4}(a). We then model each section as equivalent infinite-port quantum mechanical networks $m$=$1$ and $m$=$2$ having transmission matrices $ [T^{(m=1)}]$ and  $[T^{(m=2)}]$, respectively. Since the two networks are cascaded to each other, the overall transmission matrix $[T]$ of the entire modulating element is the product of $ [T^{(m=1)}]$ and  $[T^{(m=2)}]$ i.e. $[T]$=$[T^{(m=2)}] [T^{(m=1)}]$. It can be shown that $[T^{(m=2)}]=e^{-j2k_{op}W_o} [T^{(m=1)}]^{-1}$ {(see Appendix-C)}. Therefore, the overall transmission matrix $[T]$ of the entire modulating element of width $W$=$2W_o$ can be found out to be $[T]$=$e^{-j2k_{op}W_o} [\mathcal{I}]$. This indicates that the transmission matrix of the modulating element of width $W$=$2W_o$ is an infinite-dimensional identity matrix $[\mathcal{I}]$ multiplied by a constant $e^{-j2k_{op}W_o}$, which originates from the phase $2k_{op}W_o$ acquired by the quantum optical state as it travels the distance $2W_o$ in the waveguide. This can be visualized from Fig.~\ref{Fig4}(b). We have computed the results considering a 30GHz classical microwave reception of field-strength $|E_w|$=50V/m. Using full-wave simulation the slot-field enhancement factor $\gamma$ for slot width to be equal to 1$\mu m$ was found to be 6500. Further design details can be found in \cite{10107736}. The individual section widths were chosen as $W_o$=2.9$mm$ for 30GHz operation. From Fig.~\ref{Fig4}(b), it can be observed that the input quantum optical state with fundamental probability amplitude component ($s$=0) gets maximally split into sideband components ($s$=1) and ($s$=2) in section $m$=1. However, the maximally split probability amplitude components converge back to the initial state as the quantum optical state travels through section $m$=2. So, while the probability amplitude splitting is maximum when the modulating-element width is $W$=$W_o$, it becomes zero when the modulating-element width is $W$=$2W_o$. This shows that the electro-optic interaction-strength  cannot be maximized by monotonically increasing the modulating element width. Instead, the width has to be optimally chosen as $W_o$ (or even multiples of $W_o$) to achieve maximized electro-optic modulation-strength.

In the upcoming subsection, we show that the electro-optic modulation-depth can be further enhanced by appropriately cascading multiple modulating elements in the W-O converter.

\begin{figure*}[ht]
    \centering
\includegraphics[width=\linewidth]{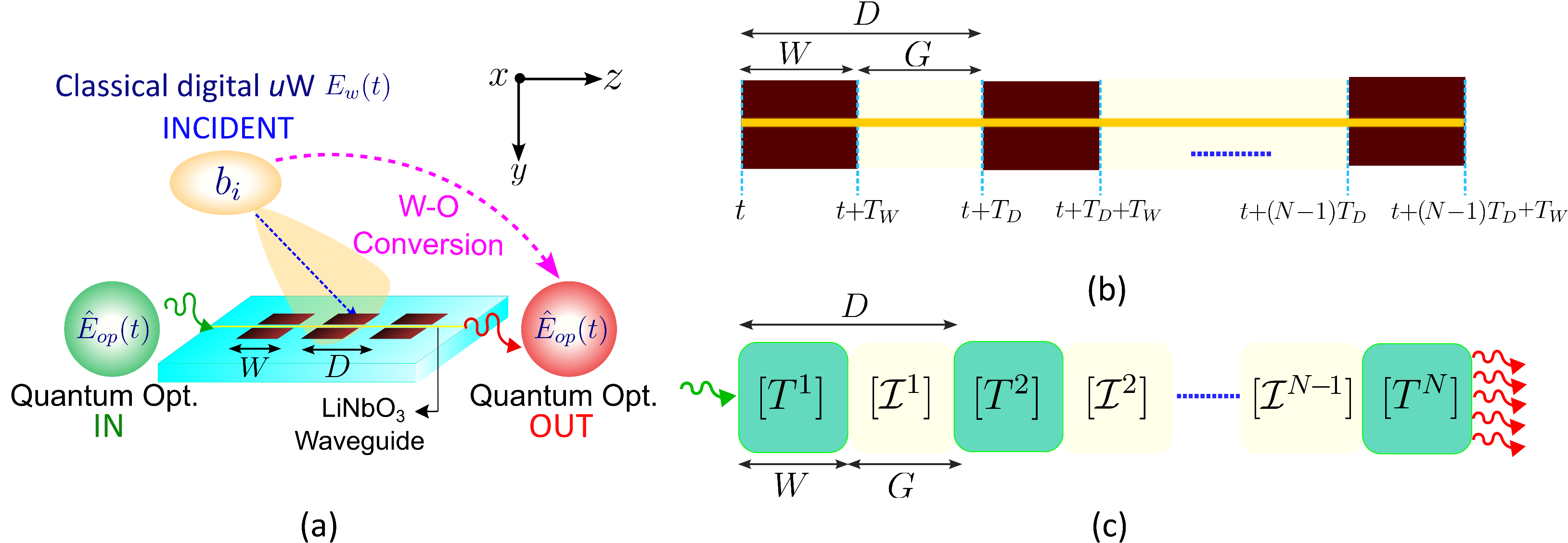}
    \caption{(a) Schematic of a W-O converter comprising of multiple cascaded modulating elements. (b) Schematic of the N-element modulating array of optimum width $W_o$ cascaded at periodicity $D$. (c) Quantum mechanical network modeling of an $N$-element modulating array.}
    \label{Fig5_2}
\end{figure*}

\subsection{Maximizing modulation-strength in an N-element converter array}

 We now consider an array of $N$ identical modulating elements of width $W$, cascaded at array periodicity $D$, as shown in Fig.~\ref{Fig5_2}(a). We then consider the quantum optical state to enter the modulating element $N$=1 at the time instant $t$, and exit the $N^{th}$ modulating element at the time instant $(t+(N-1)T_D+T_{W})$. Here, $T_{W}$=$W\sqrt{\epsilon_{op}}/c$ and $T_D$=$D\sqrt{\epsilon_{op}}/c$ is the time taken by the quantum optical state to travel distance $W$ and $D$, respectively, in the optical waveguide, as depicted in Fig.~\ref{Fig5_2}(b). Also, $G$=$(D-W)$ represents the gap between two consecutively cascaded modulating elements. We can model every modulating element in the array as an infinite-port quantum mechanical network having transmission matrix of the form shown in Eq.~(\ref{eq007}) of Appendix-D. Also, since no electro-optic interaction takes place in the optical waveguide segment of length $G$ located between the $n^{th}$ and $(n+1)^{th}$ modulating elements, it can be modeled as an infinite-port network of transmission matrix $[\mathcal{I}^{n}]$=$e^{-jk_{op}G} [\mathcal{I}]$. The overall transmission matrix $[T]$ relating the input and output probability amplitude components of the quantum optical state in the $N$-element array has been derived in Eq.~(\ref{eq32_3_x}) of Appendix-D.

The overall electro-optic modulation-depth $\delta \theta_N$ in the $N$-element array can be maximized if the time taken by the quantum optical state to travel between the respective entry points of two consecutive modulating elements is equal to the time-period of the received classical microwave E-field \cite{10107736}. This can only be achieved by optimally choosing the array periodicity as $D$=$D_o$=$(\pi c/\omega_w \sqrt{\epsilon_{op}})$, which is twice the optimum element width $W_o$. The expression of modulated-phase $\theta_i(t)$ in that case becomes the following (see derivation in Appendix-D),
\begin{equation}
  \theta_i(t)  = \delta \theta_{No}\ \cos{(\omega_w t + b_i)}
\label{eqA8_3_2y}
\end{equation}
where, the maximized electro-optic modulation-depth $\delta \theta_{No}$ can be expressed as,
\begin{equation}
\delta \theta_{No} = N \delta \theta
\label{eqA8_3_2y_2}
\end{equation}
It can be observed that the electro-optic modulation-depth gets linearly up-scaled by a factor of $N$ in an $N$-element optimally cascaded array based W-O converter. 

The total probability amplitude of the modulated quantum optical state at the array output is (see derivation in Appendix-D),
%considering only the probability amplitude component $C^{0}_1$ to be present at the input is (see derivation in Appendix-D),
%\begin{equation}
%\begin{split}
    %\Tilde{C_1} & = C^{}_1 e^{-j(\omega_{op}t+ \chi_o + \theta_i(t))} \\
    %& = C^{}_1 e^{-j(\omega_{op}t+ \chi_o)} \sum_{s=-\infty}^{s=\infty} (-1)^{s} J_s(\delta \theta_{No} )e^{-js(\omega_w t + b_i)}
%\end{split}
%\label{eq_h}
%\end{equation}
\begin{equation}
    \Tilde{C_1} = C^{}_1 e^{-j(\omega_{op}t+ \chi_o + \theta_i(t))}
\label{eq_h}
\end{equation}
where $\chi_o$=$(N-1)k_{op}D_o + k_{op}W$ is the propagation phase acquired by the modulated quantum optical state in the array. 
%Since, the electro-optic modulation-depth $\delta \theta_{N_o}$ gets upscaled in an optimally cascaded array, the probability amplitude splitting increases as the number of cascaded modulating elements are increased. Figure.~\ref{Fig6} shows the increase in probability amplitude splitting of an optical photon modulated in a modulating-element array cascaded at an array periodicity $D_o$=$2W_o$=5.8$mm$ for a received 30GHz classical microwave carrier of field-strength $|E_w|$=50V/m.

Following a similar procedure, the evolution of the probability amplitude $C_k$ of the $k^{th}$ Fock basis-state of the modulated quantum optical state can be derived to be,

\begin{equation}
 \Tilde{C_k} = C^{}_k e^{-jk (\omega_{op}t+ \chi_o + \theta_i(t))}
\label{eq_h_2}
\end{equation}
In general, a modulated quantum optical state $\ket{\psi(t)}$ composed of a superposition of Fock basis-states can be expressed as,
\begin{equation}
\ket{\psi(t)} = \sum_{k=0}^{k=\infty} C_k e^{-jk(\omega_w t  + \chi_o + \theta_i(t))}
\label{eq_h_3}
\end{equation}

\section{Quantum optical phase-space encoding with classical microwave constellation}

In this section, we extend the previously derived framework to investigate seamless phase-space encoding of quantum optical coherent-states with classical microwave constellation.

We consider an unmodulated optical coherent-state $\ket{\alpha}$ of mean complex amplitude $\alpha$ and frequency $\omega_{op}$ to be launched into a W-O converter. The W-O converter is considered to be composed of an optimally cascaded $N$-element array of modulating elements having optimum width. The received classical microwave E-field holding the $i^{th}$ digital symbol phase-modulates the optical coherent-state traveling in the W-O converter. Consequently, the modulated optical coherent-state $\ket{\alpha_i(t)}$ at the converter output can be expressed as (see Appendix-E for details), 
\begin{equation}
\ket{\alpha_i(t)} = \sum_{k=0}^{\infty} e^{-|\alpha|^2 /2}\ \frac{\big(\alpha_i(t)\big)^{k}} {\sqrt{k!}} \ket{k}
\label{eq40}
\end{equation}
where $\alpha_i(t)$=$\alpha  e^{-j(\omega_{op}t+\theta_i(t))}$ is the time-evolving mean complex amplitude (ignoring the propagation phase $\chi_o$) of the modulated coherent-state, and $\theta_i(t)$ is the mean phase encoded with the $i^{th}$ classical digital symbol. We next substitute the expression of $\theta_i(t)$ from Eq.~(\ref{eqA8_3_2y}) in the above expression of $\alpha_i(t)$. The resultant expression can then be expanded using Jacobi-Anger expansion to get,
\begin{equation}
\begin{split}
       \alpha_i(t) = \alpha \sum_{s=-\infty}^{s=\infty} (-1)^{s} J_s(\delta \theta_{N_o})\ e^{-j( \omega_{op} t + s(\omega_w t + b_i))}
\end{split}
\label{eq42}
\end{equation}
\begin{figure}[h]
    \centering
\includegraphics[width=\linewidth]{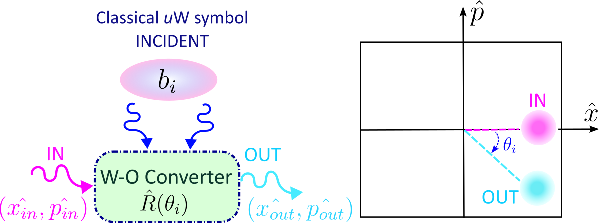}
    \caption{Unitary rotation of symbols in the encoded quantum optical phase-space due to phase-shift-keying in the classical microwave domain.}
    \label{Fig7}
\end{figure}
Considering practical values of received microwave E-field strength $|E_w|$, the corresponding values of electro-optic modulation-depth $\delta \theta_{N_o}$ are such that only a few higher-order sidebands of $\alpha_i(t)$ exist. Additionally, since $\omega_{op}>>\omega_w$, it can be argued that $\omega_{op} \pm s \omega_w \approx \omega_{op}$ in the above equation. Therefore, Eq.~(\ref{eq42}) can be approximately expressed as,
\begin{equation}
\begin{split}
      \alpha_i(t) & \approx \alpha   \sum_{s=-\infty}^{s=\infty} (-1)^{s} J_s(\delta \theta_{N_o})\ e^{-j (\omega_{op} t + s b_i)}\\
      & = \alpha e^{-j \omega_{op} t} \sum_{s=-\infty}^{s=\infty} (-1)^{s} J_s(\delta \theta_{N_o})\ e^{-j  s b_i}\\
      & =\alpha e^{-j(\omega_{op} t + \delta \theta_{No} \cos{b_i})} = \alpha e^{-j(\omega_{op} t + \theta_i)}
\end{split}
\label{eq43}
\end{equation}
where $\theta_i$ is the mean offset-phase encoded in the modulated quantum optical coherent-state, which is related to the classical microwave phase $b_i$ corresponding to the $i^{th}$ digital symbol by the following relation,
\begin{equation}
    \theta_i = \delta \theta_{No}\ \cos{b_i} = N \delta \theta\ \cos{b_i}
    \label{eq44}
\end{equation}
So, the mean location of the $i^{th}$ symbol in the modulated quantum optical $\hat{x}$-$\hat{p}$ phase-space is given by $<\hat{x}_i>$=$\Re{\{ \alpha_i \}}$ and $<\hat{p}_i>$=$\Im{\{ \alpha_i \}}$, which in is turn governed by the classical microwave phase $b_i$. Therefore, a microwave phase-shift causes the mean location of the modulated coherent-state in the quantum optical phase-space to shift. Furthermore, it can be shown that a microwave phase-shift causes a unitary rotation operation in the quantum optical phase-space. Therefore, the quadrature operators $(\hat{x_{in}}, \hat{p_{in}})$ of the unmodulated coherent-state at the converter input get transformed to the quadrature operators $(\hat{x_{out}}, \hat{p_{out}})$ at the converter output, as shown in Fig.~\ref{Fig7}. The associated matrix transformation is given as, 

\begin{equation}
    \begin{pmatrix}
         \hat{x_{out}}\\
         \hat{p_{out}}\\
        \end{pmatrix} = \hat{R}(\theta_i)  \begin{pmatrix}
           \hat{x_{in}}\\
         \hat{p_{in}}\\
        \end{pmatrix}
    \label{eq44_2}
\end{equation}
The unitary rotation operator $\hat{R}(\theta_i)$ in the above equation can be expressed as,
\begin{equation}
  \hat{R}(\theta_i) =  \begin{pmatrix}
         \cos{\theta_i} & \sin{\theta_i}\\
         \sin{\theta_i} & -\cos{\theta_i}\\
        \end{pmatrix} 
    \label{eq44_3}
\end{equation}

\begin{figure}[h]
    \centering
\includegraphics[width=\linewidth]{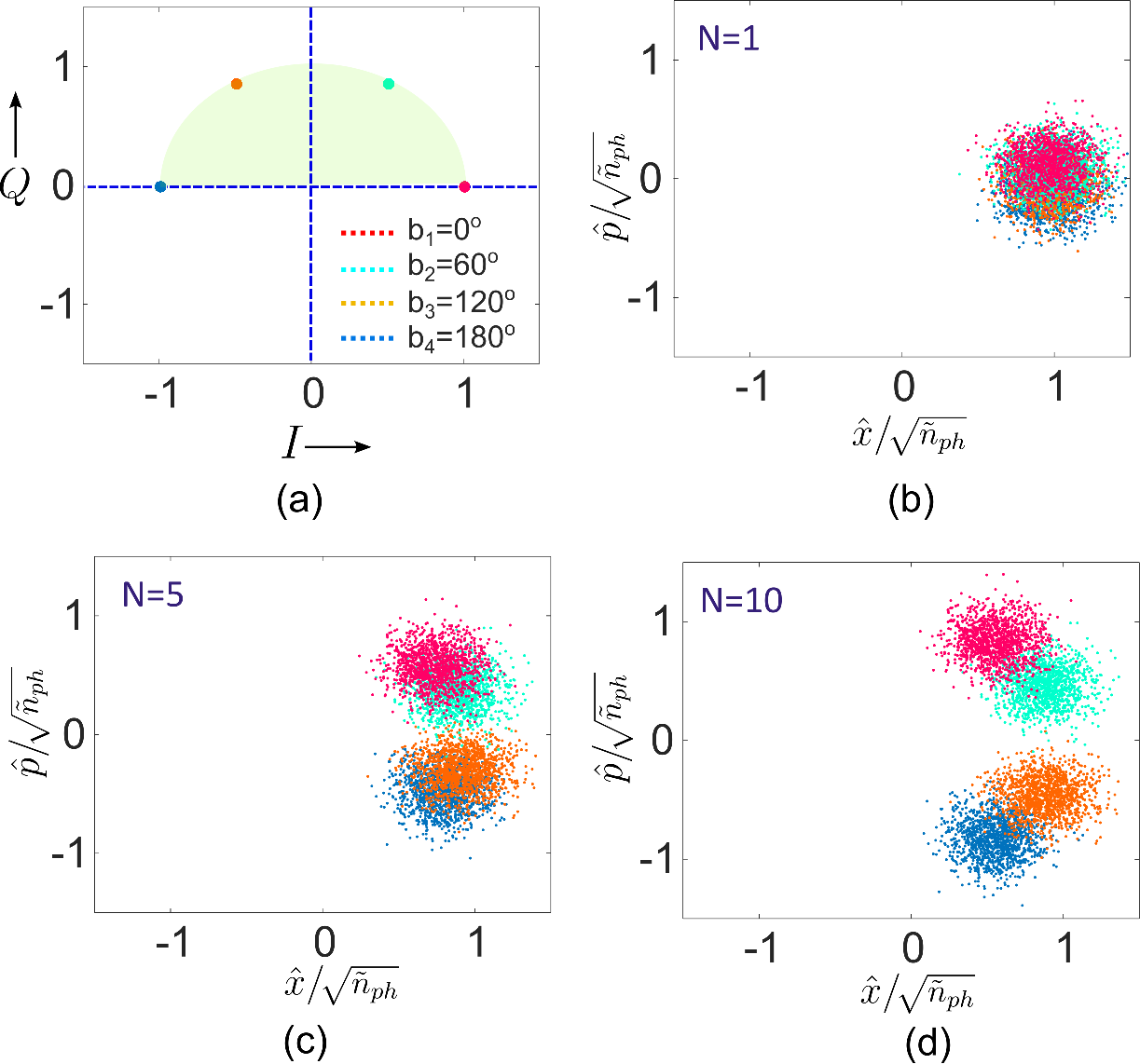}
    \caption{(a) Received 4-ary PSK constellation encapsulated in a 30GHz classical wireless microwave E-field of strength $|E_w|$=50V/m. Correspondingly encoded quantum optical phase-space of the modulated coherent-state composed of
    of mean photon numbers $\Tilde{n}_{ph}$=10 at the output of a (b) single-element, (c) 5-element, and (d) 10-element optimally cascaded array.}
    \label{Fig8}
\end{figure}

\begin{figure}[h]
    \centering
\includegraphics[width=\linewidth]{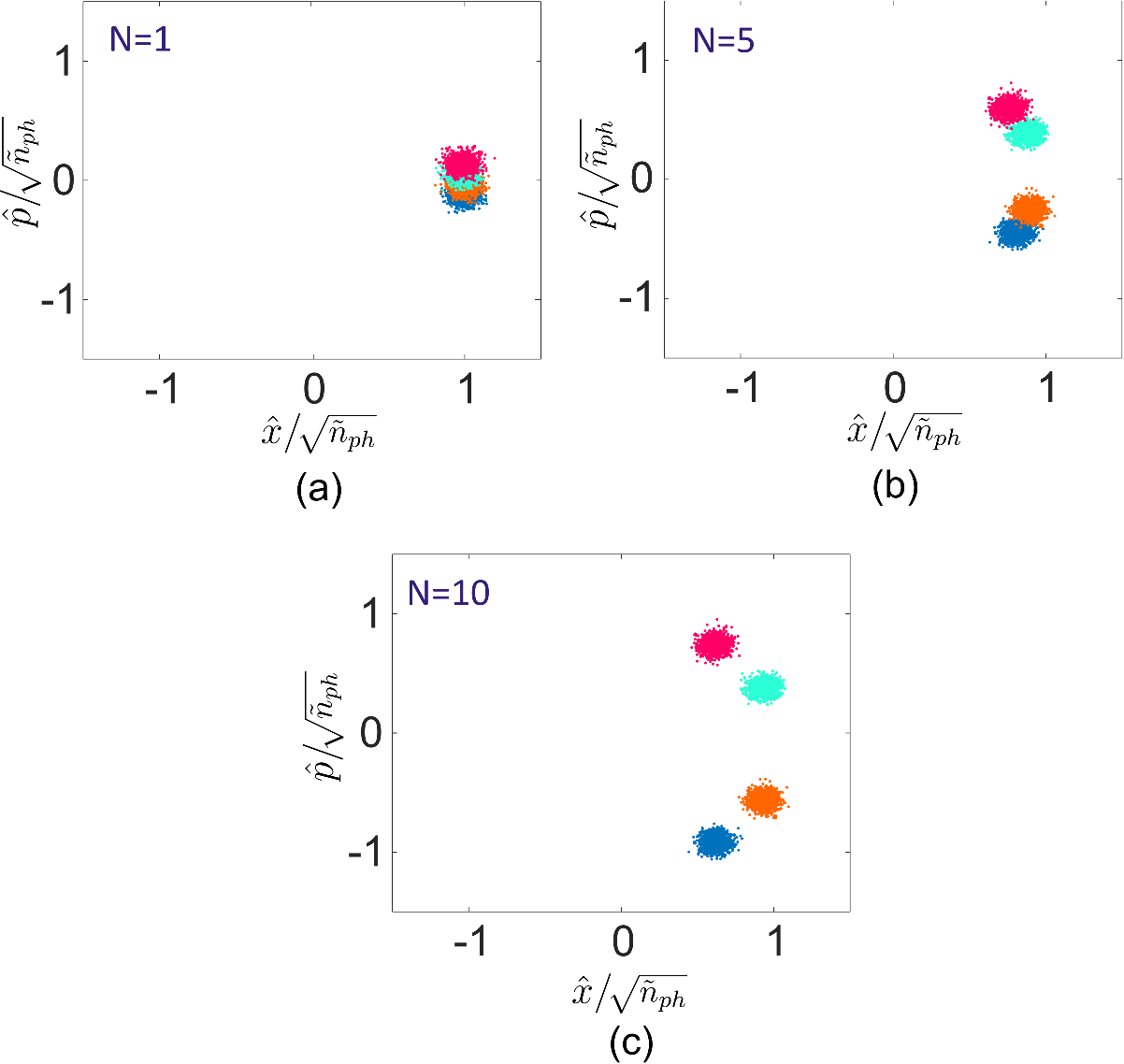}
    \caption{Encoded quantum optical phase-space of the modulated coherent-state composed of
    of mean photon numbers $\Tilde{n}_{ph}$=100 at the output of a (b) single-element, (c) 5-element, and (d) 10-element optimally cascaded array.}
    \label{Fig9}
\end{figure}

Figure \ref{Fig8}(a) shows the constellation of a modified 4-ary PSK signal encapsulated in a 30GHz classical wireless microwave carrier of field-strength $|E_w|$=50V/m. The phase-levels associated with all four possible symbols in the constellation are $b_i$ = $\{0^\circ, 60^\circ, 120^\circ, 180^\circ\}$ \cite{Ghosh_2021}. The resulting encoded phase-space of the modulated optical coherent-state at the converter output, considering the mean optical photon numbers to be $\Tilde{n}_{ph}$=$|\alpha|^{2}$=10, is shown in Fig~\ref{Fig8}(b)-(d). It can be observed that each symbol in the classical microwave constellation gets mapped to a particular location in the quantum optical phase-space. This lays the foundation for classical-to-quantum digital symbol mapping.

It must be pointed out that the jitter around the quantum optical symbols in the phase-space arises due to quadrature fluctuations. Coherent-states are minimum uncertainty states, and resemble a displaced vacuum state  \cite{furusawa2015quantum}. As a result, the quadrature variances are equal and independent of mean photon numbers contained in the coherent-state i.e. $\Delta \hat{x}$=$\Delta \hat{p}$=1/2. It can also be observed that there exists a possibility of inter-symbol overlap in the encoded quantum optical phase-space due to quadrature fluctuations. This inter-symbol overlap may render different quantum optical symbols indistinguishable, resulting in erroneous phase-space encoding. However, it can be observed from Fig~\ref{Fig8}(b)-(d) that as the number of optimally cascaded modulating elements is increased, the probability of inter-symbol interference decreases. This happens due to linear up-scaling of modulation-depth with the increase in number of optimally cascaded modulating elements (as was discussed in the previous section), causing the mean locations of different symbols in the encoded phase-space to be spaced far from each other. 

It must further be highlighted that the quadrature variances relative to the mean photon numbers i.e. $\Delta \hat{x}/\sqrt{\Tilde{n}_{ph}}$=$\Delta \hat{p}/\sqrt{\Tilde{n}_{ph}}$=$1/2\sqrt{\Tilde{n}_{ph}}$, decrease as the mean photon numbers are increased. Due to this, the inter-symbol overlap further decreases when the mean photon numbers contained in the modulated coherent-state is higher i.e. $\Tilde{n}_{ph}$=$|\alpha|^{2}$=100, as shown in Fig~\ref{Fig9}.

%This can be visualized from Fig~\ref{Fig9}, which shows that the relative jitter due to quadrature fluctuations in the quantum optical symbols decreases when the mean photon numbers are considered to be $\Tilde{n}_{ph}$=$|\alpha|^{2}$=100. 

%It must also be pointed out that there is a high probability of overlap between different symbols in the encoded quantum optical phase-space due to quantum optical shot noise. This poses the challenge of indistinguishability between different encoded symbols in the quantum optical phase-space. However, this overlap can be significantly reduced by spacing out the mean locations of each optical symbols away from each other in the phase-space. This can be achieved by increasing the number of optimally cascaded modulating elements in the W-O converter, as shown in Fig~\ref{Fig8}(c)-(d), the physical reason of which was discussed in the previous section.

%It can also be observed from Fig~\ref{Fig9} that the inter-symbol overlap in the quantum optical phase-space decreases as the mean photon numbers in the optical coherent-states decrease. This happens due to the decrease in the relative variance of the symbol locations in the quantum optical phase-space as the mean optical photon number is increased.

\section{Conclusion}
In this paper, a theoretical framework for microwave-to-optical digital information transfer is reported considering the quantum nature of light. The reported framework can be used to realize phase-space encoding of quantum optical states with classical microwave constellation. The discussed model for classical-to-quantum information transfer will form the basis for integrating digital classical microwave links with quantum optical links in the future. Even though phase-space encoding of optical coherent-states is particularly shown in this paper, the core framework is derived for a general quantum state of light. Therefore, the proposed model can be extended to study encoding of other nonclassical quantum states of light such as squeezed-states using classical microwave constellation as well, which can have a variety of interesting applications in the future. Furthermore, the reported contributions can also lay the foundation for investigating and modeling nonclassical effects in wireless-photonic receivers in next-generation communication networks \cite{10535154}.

\appendix

\section{Derivation of the electro-optic interaction Hamiltonian}
The perturbed Hamiltonian $\hat{H_I}(t)$ of the quantum optical mode due to Pockel's electro-optic effect can be derived to be,
\begin{equation}
       \hat{H}_{I}(t) =\int_{V}^{} \frac{1}{2}\ \epsilon_o\  \delta \epsilon_{op}(t)\ \hat{E}_{op}{(t)} . \hat{E}_{op}{(t)} \ dV 
\label{eqA1}
\end{equation}
where $\delta \epsilon_{op}(t)$ is the Pockel's effect-induced perturbation in the relative permittivity of the optical waveguide section channelized through the modulating element. By substituting the expression of $\delta \epsilon_{op}(t)$ from Eq.~(\ref{eq4}) and the expression of $\hat{E}_{op}(t)$ from Eq.~(\ref{eq1}) in the above integral, we get,
%\begin{widetext}
\begin{eqnarray}
       \hat{H}_{I}(t) & = \frac{\hbar \omega_{op}}{4V}\ \epsilon_{op} r_{33} E_{w}(t)\ \int_{V}^{}  \big(\hat{a}.\hat{a}\ e^{-i2(\omega_{op}t-k_{op}z)}\nonumber \\
       & -\hat{a}.\hat{a}^{\dagger} -  \hat{a}^{\dagger}. \hat{a} + \hat{a}^{\dagger}.\hat{a}^{\dagger}\  e^{i2(\omega_{op}t-k_{op}z)}    \big)\ dV
\label{eqA2}
\end{eqnarray}
%\end{widetext}
The result of integrating the terms $\hat{a}.\hat{a}\ e^{-i2(\omega_{op}t-k_{op}z)}$ and $\hat{a}^{\dagger}.\hat{a}^{\dagger}\  e^{i2(\omega_{op}t-k_{op}z)}$ becomes zero on applying periodic boundary-condition in the above integral \cite{gerry2023introductory}. Consequently, the integral of Eq.~(\ref{eqA2}) simplifies down to,
\begin{equation}
        \hat{H_I}(t) =  - \frac{\epsilon_{op} r_{33}}{2} \gamma E_w(t)\ \hbar \omega_{op}\big( \hat{a}^{\dagger}\hat{a} + \frac{1}{2} \big)
\label{eqA3}
\end{equation}
For the sake of simplicity, we consider the reference vacuum eigen energy of the quantum optical mode to be zero. In such a case, the expression of the interaction Hamiltonian $\hat{H_I}(t)$ simplifies down to,
\begin{equation}
        \hat{H_I}(t) =  - \frac{\epsilon_{op} r_{33}}{2} \gamma E_w(t)\ \hbar \omega_{op} \hat{a}^{\dagger}\hat{a}
\label{eqA3_2}
\end{equation}

\section{Derivation of time-evolution of a single optical photon state in the modulating element}

The time-evolving probability amplitude $C_1(t)$ associated with the modulated quantum optical state can be found by integrating Eq.~(\ref{eq10_2}) as follows,
\begin{equation}
   \int_{t}^{t+T_W} \frac{d C_1(t)}{C_1(t)} =-j  \int_{t}^{t+T_W} \omega_{op} \bigg(1-  \frac{\epsilon_{op} r_{33}}{2}\ \gamma E_w(t)\bigg) dt
\label{eqB_1}
\end{equation}
Substituting the expression of $E_w(t)$ from Eq.~(\ref{eq3}) and putting $T_W$=$(W\sqrt{\epsilon_{op}}/c)$ in the above integral, we solve it to get,
\begin{equation}
  C_1(t+T_W) = C_1(t) e^{-jk_{op} W} e^{-j\delta \theta \sin{(\omega_wt+\phi+b_i)}}
\label{eqB_2}
\end{equation}
where $C_1(t+T_W)$ and $C_1(t)$ are the output and input probability amplitudes at the instants $(t+T_W)$ and $t$, respectively. Also, $C_1(t)$=$C_1(0)e^{-j\omega_{op}t}$, where $C_1(0)$ is the probability amplitude at $t$=0. For the sake of notational simplicity, we represent $C_1(t+T_W)$ as $\Tilde{C_1}$, and $C_1(0)$ as $C_1$ from here on. Therefore, the time-evolving probability amplitude $C_1(t)$ can be expressed as,
\begin{equation}
\Tilde{C}_{1} = C_{1}\ e^{-j(\omega_{op}t+k_{op} W + \delta \theta \sin{(\omega_wt+\phi+b_i)})}
\label{eqB_3}
\end{equation}
Expanding Eq.~(\ref{eqB_3}) using Jacobi-Anger expansion, we get the following expression,
\begin{equation}
\Tilde{C_{1}} = C_{1} e^{-j(\omega_{op}t + k_{op}W)} \sum^{s=\infty}_{s=-\infty} j^{s} J_s(\delta \theta)\ e^{-j s({\omega_w}t +  \phi + b_i)}
\label{eqB_4}
\end{equation}
where $\delta \theta$ represents the electro-optic modulation-depth, which can be expressed as follows,
\begin{equation}
\delta \theta = -\frac{\omega_{op}\epsilon_{op} r_{33} \gamma}{\omega_w}\ \sin{\bigg(\frac{\omega_w \sqrt{\epsilon_{op}} W}{2c} \bigg)}\ |E_{w}|
\label{eqB_5}
\end{equation}

Also the offset phase term $\phi$ can be expressed as,
\begin{equation}
\phi = \frac{\omega_w \sqrt{\epsilon_{op}} W}{2c}
\label{eqB_6}
\end{equation}
It can be therefore be concluded that the initial probability amplitude of the quantum optical state splits into different sideband components upon being phase-modulated by the received classical wireless E-field. In other words, probability amplitude evolving at frequency $\omega_{op}$ splits into different sideband components evolving at frequency $\omega \pm s \omega_w$. Therefore, the output probability amplitude vector $[\Tilde{C^{s}_1}]$ and the input probability amplitude vector $[{C^{s}_1}]$ can be related by the matrix relation $[\Tilde{C^{s}_1}]$=$[T][{C^{s}_1}]$, where the transmission matrix $[T]$ can be expressed as,   

\begin{widetext}
    \begin{equation}
    [T] = e^{-jk_{op}W} \begin{bmatrix}
          \textbf{.} &  \textbf{.} & \textbf{.} & \textbf{.} & \textbf{.} & \textbf{.} & \textbf{.}\\
          \textbf{.} & \textbf{.} & \textbf{.} & \textbf{.} & \textbf{.} & \textbf{.} & \textbf{.}\\
         \textbf{.} & \textbf{.}& J_0(\delta \theta) & j^{-1} J_{-1} (\delta \theta) e^{j(\omega_{w}t +\phi + b_i)} & j^{-2} J_{-2}(\delta \theta)  e^{j2(\omega_{w}t +\phi + b_i)} & \textbf{.} & \textbf{.}\\
         \textbf{.} & \textbf{.}&  j^{} J_1(\delta \theta) e^{-j(\omega_{w}t +\phi + b_i)} & J_0(\delta \theta) & j^{-1} J_{-1}(\delta \theta)  e^{j(\omega_{w}t +\phi+ b_i)} & \textbf{.} & \textbf{.}\\
         \textbf{.} & \textbf{.} & j^{2} J_2(\delta \theta) e^{-j2(\omega_{w}t +\phi+ b_i)} & j^{1} J_1(\delta \theta) e^{-j(\omega_{w}t +\phi+ b_i)} & J_0(\delta \theta) & \textbf{.} & \textbf{.}\\
         \textbf{.} &  \textbf{.} & \textbf{.} & \textbf{.} & \textbf{.} & \textbf{.} & \textbf{.}\\
          \textbf{.} & \textbf{.} & \textbf{.} & \textbf{.} & \textbf{.} & \textbf{.} & \textbf{.}
        \end{bmatrix} 
\label{eqB_5}
\end{equation}
\end{widetext}

\section{Optimum and non-optimum modulating element width}

The transmission matrix $[T^{(m=1)}]$ of the equivalent quantum mechanical network associated with the modulating section $m$=$1$ of width $W_o$ represented in Fig.~\ref{Fig4}(a), can be derived by integrating Eq.~(\ref{eqB_1}) in the time window $t$ to $(t+T_{W_o})$. Here, $T_{W_o}$=$(W_o\sqrt{\epsilon_{op}}/c)$ is the time taken by the quantum optical state to travel distance $W_o$ in the waveguide. Following the procedure shown in the previous section, the transmission matrix $[T^{(m=1)}]$ associated with the modulating section $m$=1 in  Fig.~\ref{Fig4}(a) can be found to be,

\begin{widetext}
    \begin{equation}
    [T^{(m=1)}] =  e^{-jk_{op}W_o}\begin{bmatrix}
          \textbf{.} &  \textbf{.} & \textbf{.} & \textbf{.} & \textbf{.} & \textbf{.} & \textbf{.}\\
          \textbf{.} & \textbf{.} & \textbf{.} & \textbf{.} & \textbf{.} & \textbf{.} & \textbf{.}\\
         \textbf{.} & \textbf{.}& J_0(\delta \theta_o) & j^{-1} J_{-1} (\delta \theta_o)e^{j(\omega_{w}t +\pi/2 + b_i)} & j^{-2} J_{-2}(\delta \theta_o) e^{j2(\omega_{w}t +\pi/2 + b_i)} & \textbf{.} & \textbf{.}\\
         \textbf{.} & \textbf{.}&  j^{} J_1(\delta \theta_o) e^{-j(\omega_{w}t +\pi/2 + b_i)} & J_0(\delta \theta_o) & j^{-1} J_{-1}(\delta \theta_o) e^{j(\omega_{w}t +\pi/2+ b_i)} & \textbf{.} & \textbf{.}\\
         \textbf{.} & \textbf{.} & j^{2} J_2(\delta \theta_o) e^{-j2(\omega_{w}t +\pi/2+ b_i)} & j^{1} J_1(\delta \theta_o) e^{-j(\omega_{w}t +\pi/2+ b_i)} & J_0(\delta \theta_o) & \textbf{.} & \textbf{.}\\
         \textbf{.} &  \textbf{.} & \textbf{.} & \textbf{.} & \textbf{.} & \textbf{.} & \textbf{.}\\
          \textbf{.} & \textbf{.} & \textbf{.} & \textbf{.} & \textbf{.} & \textbf{.} & \textbf{.}
        \end{bmatrix} 
\label{eq26}
\end{equation}
\end{widetext}
where $\delta \theta_o$=$-(\omega_{op} \epsilon_{op} r_{33}/\omega_w) \gamma |E_w|$ is the electro-optic modulation depth of the modulating section $m$=1. Similarly, the  transmission matrix $[T^{(m=2)}]$ of the equivalent quantum mechanical network associated with the modulating section $m$=$2$, also having width $W_o$, as illustrated in Fig.~\ref{Fig4}(a), can be derived by integrating Eq.~(\ref{eqB_1}) in the time window $(t+T_{W_o})$ to $(t+2T_{W_o})$ to get the following,
\begin{widetext}
    \begin{equation}
    [T^{(m=2)}] =  e^{-jk_{op}W_o}\begin{bmatrix}
          \textbf{.} &  \textbf{.} & \textbf{.} & \textbf{.} & \textbf{.} & \textbf{.} & \textbf{.}\\
          \textbf{.} & \textbf{.} & \textbf{.} & \textbf{.} & \textbf{.} & \textbf{.} & \textbf{.}\\
         \textbf{.} & \textbf{.}& J_0(\delta \theta_o) & j^{-1} J_{-1} (\delta \theta_o)e^{j(\omega_{w}t +3\pi/2 + b_i)} & j^{-2} J_{-2}(\delta \theta_o) e^{j2(\omega_{w}t +3\pi/2 + b_i)} & \textbf{.} & \textbf{.}\\
         \textbf{.} & \textbf{.}&  j^{} J_1(\delta \theta_o) e^{-j(\omega_{w}t +3\pi/2 + b_i)} & J_0(\delta \theta_o) & j^{-1} J_{-1}(\delta \theta_o) e^{j(\omega_{w}t +3\pi/2+ b_i)} & \textbf{.} & \textbf{.}\\
         \textbf{.} & \textbf{.} & j^{2} J_2(\delta \theta_o) e^{-j2(\omega_{w}t +3\pi/2+ b_i)} & j^{1} J_1(\delta \theta_o) e^{-j(\omega_{w}t +3\pi/2+ b_i)} & J_0(\delta \theta_o) & \textbf{.} & \textbf{.}\\
         \textbf{.} &  \textbf{.} & \textbf{.} & \textbf{.} & \textbf{.} & \textbf{.} & \textbf{.}\\
          \textbf{.} & \textbf{.} & \textbf{.} & \textbf{.} & \textbf{.} & \textbf{.} & \textbf{.}
        \end{bmatrix} 
\label{eq28}
\end{equation}
\end{widetext}
It can be easily shown that the derived transmission matrices $[T^{(m=1)}]$ and $[T^{(m=2)}]$ are related as,
\begin{equation}
[T^{(m=1)}]=e^{-j2k_{op}W_o} [T^{(m=2)}]^{\dagger}
  \label{eq28_2}
\end{equation}
Also, since the transmission matrices $[T^{(m=1)}]$ and $[T^{(m=2)}]$ are unitary, it can be further shown that they are related to each other by the following relationship,
\begin{equation}
[T^{(m=1)}]=e^{-j2k_{op}W_o}  [T^{(m=2)}]^{-1}
  \label{eq28_3}
\end{equation}
The overall transmission matrix $[T]$ of the modulating element of width $W$=$2W_o$, which is essentially the cascade of the modulating sections $m$=1 and $m$=2, is equal to the product of the matrices $[T^{(m=1)}]$ and $[T^{(m=2)}]$. Therefore, the overall transmission matrix $[T]$ can be derived to be the following,
\begin{equation}
 [T]=[T^{(m=2)}]  [T^{(m=1)}] = e^{-j2k_{op}W_o} [\mathcal{I}]
  \label{eq30}
\end{equation}

\section{Quantum modeling of multi-element W-O converter}

The transmission matrix of the $n^{th}$ modulating element in the array represented in Fig.~\ref{Fig5_2}(b) can be derived by integrating Eq.~(\ref{eqB_1}) in the time window $(t+(n-1)T_D)$ to $(t+(n-1)T_D+T_W)$. Here, $T_D$=$D\sqrt{\epsilon_{op}}/c$ is the time taken by the quantum optical state to travel distance $D$ in the optical waveguide. The transmission matrix $[T^{n}]$ of the $n^{th}$ modulating element in the array can then be derived to be,
\begin{widetext}
    \begin{equation}
    [T^{n}] = e^{-jk_{op}W} \begin{bmatrix}
          \textbf{.} &  \textbf{.} & \textbf{.} & \textbf{.} & \textbf{.} & \textbf{.} & \textbf{.}\\
          \textbf{.} & \textbf{.} & \textbf{.} & \textbf{.} & \textbf{.} & \textbf{.} & \textbf{.}\\
         \textbf{.} & \textbf{.}& J_0(\delta \theta) & j^{-1} J_{-1} (\delta \theta) e^{j(\omega_{w}t +\phi_n + b_i)} & j^{-2} J_{-2}(\delta \theta)  e^{j2(\omega_{w}t +\phi_n + b_i)} & \textbf{.} & \textbf{.}\\
         \textbf{.} & \textbf{.}&  j^{} J_1(\delta \theta) e^{-j(\omega_{w}t +\phi_n + b_i)} & J_0(\delta \theta) & j^{-1} J_{-1}(\delta \theta)  e^{j(\omega_{w}t +\phi_n+ b_i)} & \textbf{.} & \textbf{.}\\
         \textbf{.} & \textbf{.} & j^{2} J_2(\delta \theta) e^{-j2(\omega_{w}t +\phi_n+ b_i)} & j^{1} J_1(\delta \theta) e^{-j(\omega_{w}t +\phi_n+ b_i)} & J_0(\delta \theta) & \textbf{.} & \textbf{.}\\
         \textbf{.} &  \textbf{.} & \textbf{.} & \textbf{.} & \textbf{.} & \textbf{.} & \textbf{.}\\
          \textbf{.} & \textbf{.} & \textbf{.} & \textbf{.} & \textbf{.} & \textbf{.} & \textbf{.}
        \end{bmatrix} 
\label{eq007}
\end{equation}
\end{widetext}
where the offset phase $\phi_n$ in the above matrix is,
\begin{equation}
   \phi_n  = \frac{\omega_w \sqrt{\epsilon_{op}} W}{2c} +(n-1)\frac{\omega_w \sqrt{\epsilon_{op}} D}{c}
\label{eq32}
\end{equation}
It must be pointed out that no electro-optic interaction takes place in the optical waveguide segment placed between the $n^{th}$ and the $(n+1)^{th}$ modulating element. The length of each such segment is equal to $G$=$(D-W)$, and thus can be modeled as an infinite-port network of identity transmission matrix having the following form,
\begin{equation}
   [\mathcal{I}^{n}]  = e^{-jk_{op}G} [\mathcal{I}]
\label{eq32_2}
\end{equation}
Therefore, the overall transmission matrix of the array composed of $N$ modulating elements of width $W$, cascaded at an array periodicity $D$, can be derived to be,
\begin{equation}
  [T] = [T^{N}] \prod_{n=1}^{n=(N-1)} [\mathcal{I}^{n}] [T^{n}]
\label{eq32_3}
\end{equation}

Therefore, the overall transmission matrix $[T]$ relating the input and output probability amplitude components of the quantum optical state in the $N$-element array can be derived to be,
\begin{widetext}
    \begin{equation}
    [T] = e^{-j\chi} \begin{bmatrix}
          \textbf{.} &  \textbf{.} & \textbf{.} & \textbf{.} & \textbf{.} & \textbf{.} & \textbf{.}\\
          \textbf{.} & \textbf{.} & \textbf{.} & \textbf{.} & \textbf{.} & \textbf{.} & \textbf{.}\\
         \textbf{.} & \textbf{.}& J_0(\delta\theta_{N}) & j^{-1} J_{-1} (\delta \theta_{N})e^{j(\omega_{w}t +\phi_N + b_i)} & j^{-2} J_{-2}(\delta \theta_{N}) e^{j2(\omega_{w}t +\phi_N + b_i)} & \textbf{.} & \textbf{.}\\
         \textbf{.} & \textbf{.}&  j^{} J_1(\delta \theta_{N}) e^{-j(\omega_{w}t +\phi_N + b_i)} & J_0(\delta \theta_{N}) & j^{-1} J_{-1}(\delta \theta_{N}) e^{j(\omega_{w}t +\phi_N+ b_i)} & \textbf{.} & \textbf{.}\\
         \textbf{.} & \textbf{.} & j^{2} J_2(\delta \theta_{N}) e^{-j2(\omega_{w}t +\phi_N + b_i)} & j^{1} J_1(\delta \theta_{N}) e^{-j(\omega_{w}t +\phi_N+ b_i)} & J_0(\delta \theta_{N}) & \textbf{.} & \textbf{.}\\
         \textbf{.} &  \textbf{.} & \textbf{.} & \textbf{.} & \textbf{.} & \textbf{.} & \textbf{.}\\
          \textbf{.} & \textbf{.} & \textbf{.} & \textbf{.} & \textbf{.} & \textbf{.} & \textbf{.}
        \end{bmatrix} 
\label{eq32_3_x}
\end{equation}
\end{widetext}
where $\chi$=$(N-1)k_{op}D + k_{op}W$ is the propagation phase acquired by the modulated quantum optical state in the array. It can be shown that the matrix element $T_{sp}$ belonging to the $s^{th}$ row and $p^{th}$ column of the transmission matrix $[T]$ is equal to,
%\begin{eqnarray}
     %T_{sp}  =  e^{-jk_{op}((N-1)D + W)} \sum_{x_n}^{} \big( \nonunmber \\
      %\prod_{n=1}^{n=N}  j^{x_n} J_{x_n}(\delta \theta) e^{-jx_n(\omega_w t + \phi_n+ b_i)} \big)
%\label{eqA14}
%\end{eqnarray}
\begin{equation}
   \small  T_{sp}  =  e^{-jk_{op}((N-1)D + W)} \sum_{x_n}^{} 
      \prod_{n=1}^{n=N}  j^{x_n} J_{x_n}(\delta \theta) e^{-jx_n(\omega_w t + \phi_n+ b_i)} 
\label{eqA14}
\end{equation}
where the summation $\sum_{n=1}^{N} x_n = (s-p)$ in the above expression. From the derived transmission matrix $[T]$, the $s^{th}$ sideband component of the output probability amplitude $\Tilde{C^{s}_1}$  can be found out to be,
\begin{equation}
     \Tilde{C^{s}_1} = \sum_{p=-\infty}^{p=\infty} T_{sp}\ C^{p}_{1}
\label{eqA5}
\end{equation}
where $C^{p}_{1}$ is the $p^{th}$ sideband component of the probability amplitude of the optical photon state at the input of the array. If we only consider the fundamental frequency component of the probability amplitude $C^{p=0}_1$=$C_1 e^{-j\omega_{op}t}$ of the  quantum optical state at the input of the array, the expression of $\Tilde{C^{s}_1}$ in Eq.~(\ref{eqA5}) becomes equal to the following,
\begin{equation}
     \Tilde{C^{s}_1} = T_{s0}\ C^{0}_1
\label{eqA5_2}
\end{equation}
Substituting the expression of $T_{sp}$ for $p$=0 from Eq.~(\ref{eqA14}) into the above equation, we get,
\begin{widetext}
\begin{equation}
      \Tilde{C^{s}_1}  = e^{-j(\omega_{op}t+(N-1)k_{op}D + k_{op}W)} \sum_{x_n}^{} \big( \prod_{n=1}^{n=N} j^{x_n} J_{x_n}(\delta \theta)\ e^{-jx_n(\omega_w t + \phi_n+ b_i)}\ C_1 \big)
\end{equation}
\label{eqA6}
\end{widetext}
where in this case, we have $\sum_{n=1}^{N} x_n = s$. Now, taking the summation of all the components present in the output probability amplitude vector $[\Tilde{C_1}(s)]$, we get the total probability amplitude of the modulated quantum optical state at the output of the array to be,
\begin{widetext}
\begin{equation}
\begin{split}
     \Tilde{C_1} & = \sum_{s=-\infty}^{s=\infty} \Tilde{C}_1(s)\\
     & = e^{-j(\omega_{op}t+(N-1)k_{op}D + k_{op}W)} \sum_{s=-\infty}^{s=\infty} \big( \sum_{x_n}^{}  \prod_{n=1}^{n=N} j^{x_n} J_{x_n}(\delta \theta)\ e^{-jx_n(\omega_w t + \phi_n+ b_i)}\big)\ C_1\\
     & = e^{-j(\omega_{op}t+(N-1)k_{op}D + k_{op}W)} \prod_{n=1}^{n=N} \big( \sum_{s=-\infty}^{s=\infty}  j^{s} J_{s}(\delta \theta)\ e^{-js(\omega_w t + \phi_n+ b_i)} \big)\ C_1\\
     & = e^{-j(\omega_{op}t+(N-1)k_{op}D + k_{op}W)}  \prod_{n=1}^{n=N} e^{-j \delta \theta \sin{(\omega_w t + \phi_n + b_i)}}\ C_1\\
     & = e^{-j(\omega_{op}t+(N-1)k_{op}D + k_{op}W)} \exp{\bigg( -j\sum_{n=1}^{n=N}  \delta \theta \sin{(\omega_w t + \phi_n + b_i)} \bigg)}\ C_1 \\
       & = C_1\ e^{-j(\omega_{op}t+\chi + \theta_i(t))} 
\end{split}
\label{eqA7}
\end{equation}
\end{widetext}
%where $\chi$ in the above equation is a constant phase term that is associated with the propagation phase of the modulated optical photon in the array, and can be expressed as,
%\begin{equation}
%\chi = (N-1)k_{op}D + k_{op}W
%\label{eqA13}
%\end{equation}
The summation series defining the modulated phase $\theta_i(t)$ in Eq.~(\ref{eqA7}) can be expanded and simplified in the following manner by substituting the expression of $\phi_n$ from Eq.~(\ref{eq32}) to get,
\begin{equation}
  \theta_i(t)  = \sum_{n=1}^{n=N} \delta \theta \sin{(\omega_w t+\phi_n + b_i)} = \delta \theta_N \sin{(\omega_w t  + \phi_N + b_i)}
\label{eqA8}
\end{equation}
where $\delta \theta_N$ is the overall electro-optic modulation-depth of the $N$-element array,
\begin{equation}
\delta \theta_{N}=  \delta \theta  \frac{\sin{(N \omega_w \sqrt{\epsilon_{op}}D/2)}}{\sin{(\omega_w \sqrt{\epsilon_{op}}D/2)}} 
\label{eqA8_2}
\end{equation}
Also, the offset phase term $\phi_N$ can be expressed as,
\begin{equation}
   \phi_N  =\frac{\omega_w  \sqrt{\epsilon_{op}} D}{2c} + (N-1)\frac{\omega_w \sqrt{\epsilon_{op}} D}{2c}
\label{eqA8_3}
\end{equation}
It can be shown that the modulated phase $\theta_i(t)$ can be maximized for any received symbol by choosing the array-periodicity as $D$=$2W_o$, in addition to optimally choosing the individual modulating-element width as $W_o$. The resultant expression of the modulated-phase becomes the following,
\begin{equation}
  \theta_i(t)  = \delta \theta_{No}\ \cos{(\omega_w t + b_i)}
\label{eqA8_3_2}
\end{equation}
where, the maximized electro-optical modulation-depth $\delta \theta_{No}$ can be expressed as,
\begin{equation}
\delta \theta_{No} = N \delta \theta
\label{eqA8_3_2}
\end{equation}

\section{Derivation of phase encoding of optical coherent-states}
An optical coherent-state $\ket{\alpha}$ can be defined as a superposition of Fock basis-states obeying Poissonian distribution, and can be expressed as follows \cite{furusawa2015quantum},
\begin{equation}
\ket{\alpha} = \sum_{k=0}^{\infty} C_k \ket{k} = \sum_{k=0}^{\infty} e^{-|\alpha|^2 /2}\ \frac{(\alpha )^{k}} {\sqrt{k!}} \ket{k}
\label{eq_AE1}
\end{equation}
where $\alpha$ is the mean complex amplitude of the E-field associated with the optical coherent-state \cite{asavanant2022optical}. We then consider the case where the optical coherent-state is subjected to phase-modulation by the $i^{th}$ digital symbol encapsulated in the received microwave E-field in an optimally cascaded N-element W-O converter. From Eq.~(\ref{eq_h_3}), the time-evolution of the modulated optical coherent-state $\ket{\alpha_i(t)}$ at the converter output can be derived to be,
\begin{equation}
\begin{split}
    \ket{\alpha_i (t)} & = \sum_{k=0}^{\infty} C_k e^{-jk(\omega_{op} t  + \chi_o + \theta_i(t))} \ket{k} \\
    & = \sum_{k=0}^{\infty} e^{-|\alpha|^2 /2}\ \frac{(\alpha )^{k}} {\sqrt{k!}} e^{-jk(\omega_{op} t  + \chi_o + \theta_i(t))} \ket{k} \\
    & =  \sum_{k=0}^{\infty} e^{-|\alpha|^2 /2}\ \frac{\big(\alpha e^{-jk(\omega_w t  + \chi_o + \theta_i(t))}\big)^{k}} {\sqrt{k!}} \ket{k} \\
    & = \sum_{k=0}^{\infty} e^{-|\alpha|^2 /2}\ \frac{\big(\alpha_i(t) \big)^{k}} {\sqrt{k!}} \ket{k} 
\end{split}
\label{eq_AE2}
\end{equation}
where $\alpha_i(t)$=$\alpha e^{-jk(\omega_w t  + \chi_o + \theta_i(t))}$ is the time-evolving mean complex amplitude of the modulated optical coherent-state.

%\bibliography{apssamp}
%\bibliographystyle{ieeetr}
%\bibliography{Bibliography}

\begin{thebibliography}{0}%
\makeatletter
\providecommand \@ifxundefined [1]{%
 \@ifx{#1\undefined}
}%
\providecommand \@ifnum [1]{%
 \ifnum #1\expandafter \@firstoftwo
 \else \expandafter \@secondoftwo
 \fi
}%
\providecommand \@ifx [1]{%
 \ifx #1\expandafter \@firstoftwo
 \else \expandafter \@secondoftwo
 \fi
}%
\providecommand \natexlab [1]{#1}%
\providecommand \enquote  [1]{``#1''}%
\providecommand \bibnamefont  [1]{#1}%
\providecommand \bibfnamefont [1]{#1}%
\providecommand \citenamefont [1]{#1}%
\providecommand \href@noop [0]{\@secondoftwo}%
\providecommand \href [0]{\begingroup \@sanitize@url \@href}%
\providecommand \@href[1]{\@@startlink{#1}\@@href}%
\providecommand \@@href[1]{\endgroup#1\@@endlink}%
\providecommand \@sanitize@url [0]{\catcode `\\12\catcode `\$12\catcode
  `\&12\catcode `\#12\catcode `\^12\catcode `\_12\catcode `\%12\relax}%
\providecommand \@@startlink[1]{}%
\providecommand \@@endlink[0]{}%
\providecommand \url  [0]{\begingroup\@sanitize@url \@url }%
\providecommand \@url [1]{\endgroup\@href {#1}{\urlprefix }}%
\providecommand \urlprefix  [0]{URL }%
\providecommand \Eprint [0]{\href }%
\providecommand \doibase [0]{https://doi.org/}%
\providecommand \selectlanguage [0]{\@gobble}%
\providecommand \bibinfo  [0]{\@secondoftwo}%
\providecommand \bibfield  [0]{\@secondoftwo}%
\providecommand \translation [1]{[#1]}%
\providecommand \BibitemOpen [0]{}%
\providecommand \bibitemStop [0]{}%
\providecommand \bibitemNoStop [0]{.\EOS\space}%
\providecommand \EOS [0]{\spacefactor3000\relax}%
\providecommand \BibitemShut  [1]{\csname bibitem#1\endcsname}%
\let\auto@bib@innerbib\@empty
%</preamble>
\end{thebibliography}%


\begin{thebibliography}{34}%
\makeatletter
\providecommand \@ifxundefined [1]{%
 \@ifx{#1\undefined}
}%
\providecommand \@ifnum [1]{%
 \ifnum #1\expandafter \@firstoftwo
 \else \expandafter \@secondoftwo
 \fi
}%
\providecommand \@ifx [1]{%
 \ifx #1\expandafter \@firstoftwo
 \else \expandafter \@secondoftwo
 \fi
}%
\providecommand \natexlab [1]{#1}%
\providecommand \enquote  [1]{``#1''}%
\providecommand \bibnamefont  [1]{#1}%
\providecommand \bibfnamefont [1]{#1}%
\providecommand \citenamefont [1]{#1}%
\providecommand \href@noop [0]{\@secondoftwo}%
\providecommand \href [0]{\begingroup \@sanitize@url \@href}%
\providecommand \@href[1]{\@@startlink{#1}\@@href}%
\providecommand \@@href[1]{\endgroup#1\@@endlink}%
\providecommand \@sanitize@url [0]{\catcode `\\12\catcode `\$12\catcode
  `\&12\catcode `\#12\catcode `\^12\catcode `\_12\catcode `\%12\relax}%
\providecommand \@@startlink[1]{}%
\providecommand \@@endlink[0]{}%
\providecommand \url  [0]{\begingroup\@sanitize@url \@url }%
\providecommand \@url [1]{\endgroup\@href {#1}{\urlprefix }}%
\providecommand \urlprefix  [0]{URL }%
\providecommand \Eprint [0]{\href }%
\providecommand \doibase [0]{https://doi.org/}%
\providecommand \selectlanguage [0]{\@gobble}%
\providecommand \bibinfo  [0]{\@secondoftwo}%
\providecommand \bibfield  [0]{\@secondoftwo}%
\providecommand \translation [1]{[#1]}%
\providecommand \BibitemOpen [0]{}%
\providecommand \bibitemStop [0]{}%
\providecommand \bibitemNoStop [0]{.\EOS\space}%
\providecommand \EOS [0]{\spacefactor3000\relax}%
\providecommand \BibitemShut  [1]{\csname bibitem#1\endcsname}%
\let\auto@bib@innerbib\@empty
%</preamble>
\bibitem [{\citenamefont {Zhu}\ \emph {et~al.}(2024)\citenamefont {Zhu},
  \citenamefont {Yoshida}, \citenamefont {Akahane},\ and\ \citenamefont
  {Kitayama}}]{zhu2024high}%
  \BibitemOpen
  \bibfield  {author} {\bibinfo {author} {\bibfnamefont {P.}~\bibnamefont
  {Zhu}}, \bibinfo {author} {\bibfnamefont {Y.}~\bibnamefont {Yoshida}},
  \bibinfo {author} {\bibfnamefont {K.}~\bibnamefont {Akahane}},\ and\ \bibinfo
  {author} {\bibfnamefont {K.-i.}\ \bibnamefont {Kitayama}},\ }\bibfield
  {title} {\bibinfo {title} {High-speed reach-extended im-dd system with
  low-complexity dsp for 6g fronthaul},\ }\href@noop {} {\bibfield  {journal}
  {\bibinfo  {journal} {Journal of Optical Communications and Networking}\
  }\textbf {\bibinfo {volume} {16}},\ \bibinfo {pages} {A24} (\bibinfo {year}
  {2024})}\BibitemShut {NoStop}%
\bibitem [{\citenamefont {Filgueiras}\ \emph {et~al.}(2023)\citenamefont
  {Filgueiras}, \citenamefont {Lima}, \citenamefont {Cunha}, \citenamefont
  {Lopes}, \citenamefont {De~Souza}, \citenamefont {Borges}, \citenamefont
  {Pereira}, \citenamefont {Brand{\~a}o}, \citenamefont {Andrade},
  \citenamefont {Alexandre} \emph {et~al.}}]{filgueiras2023wireless}%
  \BibitemOpen
  \bibfield  {author} {\bibinfo {author} {\bibfnamefont {H.~R.~D.}\
  \bibnamefont {Filgueiras}}, \bibinfo {author} {\bibfnamefont {E.~S.}\
  \bibnamefont {Lima}}, \bibinfo {author} {\bibfnamefont {M.~S.~B.}\
  \bibnamefont {Cunha}}, \bibinfo {author} {\bibfnamefont {C.~H. D.~S.}\
  \bibnamefont {Lopes}}, \bibinfo {author} {\bibfnamefont {L.~C.}\ \bibnamefont
  {De~Souza}}, \bibinfo {author} {\bibfnamefont {R.~M.}\ \bibnamefont
  {Borges}}, \bibinfo {author} {\bibfnamefont {L.~A.~M.}\ \bibnamefont
  {Pereira}}, \bibinfo {author} {\bibfnamefont {T.~H.}\ \bibnamefont
  {Brand{\~a}o}}, \bibinfo {author} {\bibfnamefont {T.~P.~V.}\ \bibnamefont
  {Andrade}}, \bibinfo {author} {\bibfnamefont {L.~C.}\ \bibnamefont
  {Alexandre}}, \emph {et~al.},\ }\bibfield  {title} {\bibinfo {title}
  {Wireless and optical convergent access technologies toward 6g},\ }\href@noop
  {} {\bibfield  {journal} {\bibinfo  {journal} {IEEE Access}\ }\textbf
  {\bibinfo {volume} {11}},\ \bibinfo {pages} {9232} (\bibinfo {year}
  {2023})}\BibitemShut {NoStop}%
\bibitem [{\citenamefont {L{\'o}pez-Cardona}\ \emph {et~al.}(2021)\citenamefont
  {L{\'o}pez-Cardona}, \citenamefont {Rommel}, \citenamefont {Grivas},
  \citenamefont {Montero}, \citenamefont {Dubov}, \citenamefont {Kritharidis},
  \citenamefont {Tafur-Monroy},\ and\ \citenamefont
  {Vazquez}}]{lopez2021power}%
  \BibitemOpen
  \bibfield  {author} {\bibinfo {author} {\bibfnamefont {J.~D.}\ \bibnamefont
  {L{\'o}pez-Cardona}}, \bibinfo {author} {\bibfnamefont {S.}~\bibnamefont
  {Rommel}}, \bibinfo {author} {\bibfnamefont {E.}~\bibnamefont {Grivas}},
  \bibinfo {author} {\bibfnamefont {D.~S.}\ \bibnamefont {Montero}}, \bibinfo
  {author} {\bibfnamefont {M.}~\bibnamefont {Dubov}}, \bibinfo {author}
  {\bibfnamefont {D.}~\bibnamefont {Kritharidis}}, \bibinfo {author}
  {\bibfnamefont {I.}~\bibnamefont {Tafur-Monroy}},\ and\ \bibinfo {author}
  {\bibfnamefont {C.}~\bibnamefont {Vazquez}},\ }\bibfield  {title} {\bibinfo
  {title} {Power-over-fiber in a 10 km long multicore fiber link within a 5g
  fronthaul scenario},\ }\href@noop {} {\bibfield  {journal} {\bibinfo
  {journal} {Optics Letters}\ }\textbf {\bibinfo {volume} {46}},\ \bibinfo
  {pages} {5348} (\bibinfo {year} {2021})}\BibitemShut {NoStop}%
\bibitem [{\citenamefont {Lim}\ \emph {et~al.}(2019)\citenamefont {Lim},
  \citenamefont {Tian}, \citenamefont {Ranaweera}, \citenamefont {Nirmalathas},
  \citenamefont {Wong},\ and\ \citenamefont {Lee}}]{lim2019evolution}%
  \BibitemOpen
  \bibfield  {author} {\bibinfo {author} {\bibfnamefont {C.}~\bibnamefont
  {Lim}}, \bibinfo {author} {\bibfnamefont {Y.}~\bibnamefont {Tian}}, \bibinfo
  {author} {\bibfnamefont {C.}~\bibnamefont {Ranaweera}}, \bibinfo {author}
  {\bibfnamefont {T.~A.}\ \bibnamefont {Nirmalathas}}, \bibinfo {author}
  {\bibfnamefont {E.}~\bibnamefont {Wong}},\ and\ \bibinfo {author}
  {\bibfnamefont {K.-L.}\ \bibnamefont {Lee}},\ }\bibfield  {title} {\bibinfo
  {title} {Evolution of radio-over-fiber technology},\ }\href@noop {}
  {\bibfield  {journal} {\bibinfo  {journal} {Journal of Lightwave Technology}\
  }\textbf {\bibinfo {volume} {37}},\ \bibinfo {pages} {1647} (\bibinfo {year}
  {2019})}\BibitemShut {NoStop}%
\bibitem [{\citenamefont {Jia}\ \emph {et~al.}(2018)\citenamefont {Jia},
  \citenamefont {Pang}, \citenamefont {Ozolins}, \citenamefont {Yu},
  \citenamefont {Hu}, \citenamefont {Yu}, \citenamefont {Guan}, \citenamefont
  {Da~Ros}, \citenamefont {Popov}, \citenamefont {Jacobsen} \emph
  {et~al.}}]{jia20180}%
  \BibitemOpen
  \bibfield  {author} {\bibinfo {author} {\bibfnamefont {S.}~\bibnamefont
  {Jia}}, \bibinfo {author} {\bibfnamefont {X.}~\bibnamefont {Pang}}, \bibinfo
  {author} {\bibfnamefont {O.}~\bibnamefont {Ozolins}}, \bibinfo {author}
  {\bibfnamefont {X.}~\bibnamefont {Yu}}, \bibinfo {author} {\bibfnamefont
  {H.}~\bibnamefont {Hu}}, \bibinfo {author} {\bibfnamefont {J.}~\bibnamefont
  {Yu}}, \bibinfo {author} {\bibfnamefont {P.}~\bibnamefont {Guan}}, \bibinfo
  {author} {\bibfnamefont {F.}~\bibnamefont {Da~Ros}}, \bibinfo {author}
  {\bibfnamefont {S.}~\bibnamefont {Popov}}, \bibinfo {author} {\bibfnamefont
  {G.}~\bibnamefont {Jacobsen}}, \emph {et~al.},\ }\bibfield  {title} {\bibinfo
  {title} {0.4 thz photonic-wireless link with 106 gb/s single channel
  bitrate},\ }\href@noop {} {\bibfield  {journal} {\bibinfo  {journal} {Journal
  of Lightwave Technology}\ }\textbf {\bibinfo {volume} {36}},\ \bibinfo
  {pages} {610} (\bibinfo {year} {2018})}\BibitemShut {NoStop}%
\bibitem [{\citenamefont {N~Fernando}(2023)}]{n2023radio}%
  \BibitemOpen
  \bibfield  {author} {\bibinfo {author} {\bibfnamefont {X.}~\bibnamefont
  {N~Fernando}},\ }\href@noop {} {\bibinfo {title} {Radio over fiber for
  wireless communications from fundamentals to advanced topics}} (\bibinfo
  {year} {2023})\BibitemShut {NoStop}%
\bibitem [{\citenamefont {Akyildiz}\ \emph {et~al.}(2022)\citenamefont
  {Akyildiz}, \citenamefont {Han}, \citenamefont {Hu}, \citenamefont {Nie},\
  and\ \citenamefont {Jornet}}]{akyildiz2022terahertz}%
  \BibitemOpen
  \bibfield  {author} {\bibinfo {author} {\bibfnamefont {I.~F.}\ \bibnamefont
  {Akyildiz}}, \bibinfo {author} {\bibfnamefont {C.}~\bibnamefont {Han}},
  \bibinfo {author} {\bibfnamefont {Z.}~\bibnamefont {Hu}}, \bibinfo {author}
  {\bibfnamefont {S.}~\bibnamefont {Nie}},\ and\ \bibinfo {author}
  {\bibfnamefont {J.~M.}\ \bibnamefont {Jornet}},\ }\bibfield  {title}
  {\bibinfo {title} {Terahertz band communication: An old problem revisited and
  research directions for the next decade},\ }\href@noop {} {\bibfield
  {journal} {\bibinfo  {journal} {IEEE Transactions on Communications}\
  }\textbf {\bibinfo {volume} {70}},\ \bibinfo {pages} {4250} (\bibinfo {year}
  {2022})}\BibitemShut {NoStop}%
\bibitem [{\citenamefont {Burla}\ \emph {et~al.}(2019)\citenamefont {Burla},
  \citenamefont {Hoessbacher}, \citenamefont {Heni}, \citenamefont {Haffner},
  \citenamefont {Fedoryshyn}, \citenamefont {Werner}, \citenamefont {Watanabe},
  \citenamefont {Massler}, \citenamefont {Elder}, \citenamefont {Dalton} \emph
  {et~al.}}]{burla2019500}%
  \BibitemOpen
  \bibfield  {author} {\bibinfo {author} {\bibfnamefont {M.}~\bibnamefont
  {Burla}}, \bibinfo {author} {\bibfnamefont {C.}~\bibnamefont {Hoessbacher}},
  \bibinfo {author} {\bibfnamefont {W.}~\bibnamefont {Heni}}, \bibinfo {author}
  {\bibfnamefont {C.}~\bibnamefont {Haffner}}, \bibinfo {author} {\bibfnamefont
  {Y.}~\bibnamefont {Fedoryshyn}}, \bibinfo {author} {\bibfnamefont
  {D.}~\bibnamefont {Werner}}, \bibinfo {author} {\bibfnamefont
  {T.}~\bibnamefont {Watanabe}}, \bibinfo {author} {\bibfnamefont
  {H.}~\bibnamefont {Massler}}, \bibinfo {author} {\bibfnamefont {D.~L.}\
  \bibnamefont {Elder}}, \bibinfo {author} {\bibfnamefont {L.~R.}\ \bibnamefont
  {Dalton}}, \emph {et~al.},\ }\bibfield  {title} {\bibinfo {title} {500 ghz
  plasmonic mach-zehnder modulator enabling sub-thz microwave photonics},\
  }\href@noop {} {\bibfield  {journal} {\bibinfo  {journal} {Apl Photonics}\
  }\textbf {\bibinfo {volume} {4}} (\bibinfo {year} {2019})}\BibitemShut
  {NoStop}%
\bibitem [{\citenamefont {Harter}\ \emph {et~al.}(2019)\citenamefont {Harter},
  \citenamefont {Ummethala}, \citenamefont {Blaicher}, \citenamefont
  {Muehlbrandt}, \citenamefont {Wolf}, \citenamefont {Weber}, \citenamefont
  {Adib}, \citenamefont {Kemal}, \citenamefont {Merboldt}, \citenamefont {Boes}
  \emph {et~al.}}]{harter2019wireless}%
  \BibitemOpen
  \bibfield  {author} {\bibinfo {author} {\bibfnamefont {T.}~\bibnamefont
  {Harter}}, \bibinfo {author} {\bibfnamefont {S.}~\bibnamefont {Ummethala}},
  \bibinfo {author} {\bibfnamefont {M.}~\bibnamefont {Blaicher}}, \bibinfo
  {author} {\bibfnamefont {S.}~\bibnamefont {Muehlbrandt}}, \bibinfo {author}
  {\bibfnamefont {S.}~\bibnamefont {Wolf}}, \bibinfo {author} {\bibfnamefont
  {M.}~\bibnamefont {Weber}}, \bibinfo {author} {\bibfnamefont {M.~M.~H.}\
  \bibnamefont {Adib}}, \bibinfo {author} {\bibfnamefont {J.~N.}\ \bibnamefont
  {Kemal}}, \bibinfo {author} {\bibfnamefont {M.}~\bibnamefont {Merboldt}},
  \bibinfo {author} {\bibfnamefont {F.}~\bibnamefont {Boes}}, \emph {et~al.},\
  }\bibfield  {title} {\bibinfo {title} {Wireless thz link with optoelectronic
  transmitter and receiver},\ }\href@noop {} {\bibfield  {journal} {\bibinfo
  {journal} {Optica}\ }\textbf {\bibinfo {volume} {6}},\ \bibinfo {pages}
  {1063} (\bibinfo {year} {2019})}\BibitemShut {NoStop}%
\bibitem [{\citenamefont {Nagatsuma}\ \emph {et~al.}(2016)\citenamefont
  {Nagatsuma}, \citenamefont {Ducournau},\ and\ \citenamefont
  {Renaud}}]{nagatsuma2016advances}%
  \BibitemOpen
  \bibfield  {author} {\bibinfo {author} {\bibfnamefont {T.}~\bibnamefont
  {Nagatsuma}}, \bibinfo {author} {\bibfnamefont {G.}~\bibnamefont
  {Ducournau}},\ and\ \bibinfo {author} {\bibfnamefont {C.~C.}\ \bibnamefont
  {Renaud}},\ }\bibfield  {title} {\bibinfo {title} {Advances in terahertz
  communications accelerated by photonics},\ }\href@noop {} {\bibfield
  {journal} {\bibinfo  {journal} {Nature Photonics}\ }\textbf {\bibinfo
  {volume} {10}},\ \bibinfo {pages} {371} (\bibinfo {year} {2016})}\BibitemShut
  {NoStop}%
\bibitem [{\citenamefont {Ummethala}\ \emph {et~al.}(2019)\citenamefont
  {Ummethala}, \citenamefont {Harter}, \citenamefont {Koehnle}, \citenamefont
  {Li}, \citenamefont {Muehlbrandt}, \citenamefont {Kutuvantavida},
  \citenamefont {Kemal}, \citenamefont {Marin-Palomo}, \citenamefont
  {Schaefer}, \citenamefont {Tessmann} \emph {et~al.}}]{ummethala2019thz}%
  \BibitemOpen
  \bibfield  {author} {\bibinfo {author} {\bibfnamefont {S.}~\bibnamefont
  {Ummethala}}, \bibinfo {author} {\bibfnamefont {T.}~\bibnamefont {Harter}},
  \bibinfo {author} {\bibfnamefont {K.}~\bibnamefont {Koehnle}}, \bibinfo
  {author} {\bibfnamefont {Z.}~\bibnamefont {Li}}, \bibinfo {author}
  {\bibfnamefont {S.}~\bibnamefont {Muehlbrandt}}, \bibinfo {author}
  {\bibfnamefont {Y.}~\bibnamefont {Kutuvantavida}}, \bibinfo {author}
  {\bibfnamefont {J.}~\bibnamefont {Kemal}}, \bibinfo {author} {\bibfnamefont
  {P.}~\bibnamefont {Marin-Palomo}}, \bibinfo {author} {\bibfnamefont
  {J.}~\bibnamefont {Schaefer}}, \bibinfo {author} {\bibfnamefont
  {A.}~\bibnamefont {Tessmann}}, \emph {et~al.},\ }\bibfield  {title} {\bibinfo
  {title} {Thz-to-optical conversion in wireless communications using an
  ultra-broadband plasmonic modulator},\ }\href@noop {} {\bibfield  {journal}
  {\bibinfo  {journal} {Nature photonics}\ }\textbf {\bibinfo {volume} {13}},\
  \bibinfo {pages} {519} (\bibinfo {year} {2019})}\BibitemShut {NoStop}%
\bibitem [{\citenamefont {Li}\ \emph {et~al.}(2014)\citenamefont {Li},
  \citenamefont {Yu}, \citenamefont {Xiao},\ and\ \citenamefont
  {Xu}}]{li2014fiber}%
  \BibitemOpen
  \bibfield  {author} {\bibinfo {author} {\bibfnamefont {X.}~\bibnamefont
  {Li}}, \bibinfo {author} {\bibfnamefont {J.}~\bibnamefont {Yu}}, \bibinfo
  {author} {\bibfnamefont {J.}~\bibnamefont {Xiao}},\ and\ \bibinfo {author}
  {\bibfnamefont {Y.}~\bibnamefont {Xu}},\ }\bibfield  {title} {\bibinfo
  {title} {Fiber-wireless-fiber link for 128-gb/s pdm-16qam signal transmission
  at$\backslash$(w$\backslash$)-band},\ }\href@noop {} {\bibfield  {journal}
  {\bibinfo  {journal} {IEEE Photonics Technology Letters}\ }\textbf {\bibinfo
  {volume} {26}},\ \bibinfo {pages} {1948} (\bibinfo {year}
  {2014})}\BibitemShut {NoStop}%
\bibitem [{\citenamefont {Zhang}\ \emph {et~al.}(2022)\citenamefont {Zhang},
  \citenamefont {Zhu}, \citenamefont {Lei}, \citenamefont {Hua}, \citenamefont
  {Cai}, \citenamefont {Zou}, \citenamefont {Tian}, \citenamefont {Li},
  \citenamefont {Wang}, \citenamefont {Huang} \emph {et~al.}}]{zhang2022real}%
  \BibitemOpen
  \bibfield  {author} {\bibinfo {author} {\bibfnamefont {J.}~\bibnamefont
  {Zhang}}, \bibinfo {author} {\bibfnamefont {M.}~\bibnamefont {Zhu}}, \bibinfo
  {author} {\bibfnamefont {M.}~\bibnamefont {Lei}}, \bibinfo {author}
  {\bibfnamefont {B.}~\bibnamefont {Hua}}, \bibinfo {author} {\bibfnamefont
  {Y.}~\bibnamefont {Cai}}, \bibinfo {author} {\bibfnamefont {Y.}~\bibnamefont
  {Zou}}, \bibinfo {author} {\bibfnamefont {L.}~\bibnamefont {Tian}}, \bibinfo
  {author} {\bibfnamefont {A.}~\bibnamefont {Li}}, \bibinfo {author}
  {\bibfnamefont {Y.}~\bibnamefont {Wang}}, \bibinfo {author} {\bibfnamefont
  {Y.}~\bibnamefont {Huang}}, \emph {et~al.},\ }\bibfield  {title} {\bibinfo
  {title} {Real-time demonstration of 103.125-gbps fiber--thz--fiber 2$\times$
  2 mimo transparent transmission at 360--430 ghz based on photonics},\
  }\href@noop {} {\bibfield  {journal} {\bibinfo  {journal} {Optics letters}\
  }\textbf {\bibinfo {volume} {47}},\ \bibinfo {pages} {1214} (\bibinfo {year}
  {2022})}\BibitemShut {NoStop}%
\bibitem [{\citenamefont {Cai}\ \emph {et~al.}(2023)\citenamefont {Cai},
  \citenamefont {Zhu}, \citenamefont {Zhang}, \citenamefont {Lei},
  \citenamefont {Hua}, \citenamefont {Zou}, \citenamefont {Luo}, \citenamefont
  {Xiang}, \citenamefont {Tian}, \citenamefont {Ding} \emph
  {et~al.}}]{cai2023real}%
  \BibitemOpen
  \bibfield  {author} {\bibinfo {author} {\bibfnamefont {Y.}~\bibnamefont
  {Cai}}, \bibinfo {author} {\bibfnamefont {M.}~\bibnamefont {Zhu}}, \bibinfo
  {author} {\bibfnamefont {J.}~\bibnamefont {Zhang}}, \bibinfo {author}
  {\bibfnamefont {M.}~\bibnamefont {Lei}}, \bibinfo {author} {\bibfnamefont
  {B.}~\bibnamefont {Hua}}, \bibinfo {author} {\bibfnamefont {Y.}~\bibnamefont
  {Zou}}, \bibinfo {author} {\bibfnamefont {W.}~\bibnamefont {Luo}}, \bibinfo
  {author} {\bibfnamefont {S.}~\bibnamefont {Xiang}}, \bibinfo {author}
  {\bibfnamefont {L.}~\bibnamefont {Tian}}, \bibinfo {author} {\bibfnamefont
  {J.}~\bibnamefont {Ding}}, \emph {et~al.},\ }\bibfield  {title} {\bibinfo
  {title} {Real-time 100-gbe fiber-wireless seamless integration system using
  an electromagnetic dual-polarized single-input single-output wireless link at
  the w band},\ }\href@noop {} {\bibfield  {journal} {\bibinfo  {journal}
  {Optics Letters}\ }\textbf {\bibinfo {volume} {48}},\ \bibinfo {pages} {928}
  (\bibinfo {year} {2023})}\BibitemShut {NoStop}%
\bibitem [{\citenamefont {Horst}\ \emph {et~al.}(2022)\citenamefont {Horst},
  \citenamefont {Blatter}, \citenamefont {Kulmer}, \citenamefont {Bitachon},
  \citenamefont {Baeuerle}, \citenamefont {Destraz}, \citenamefont {Heni},
  \citenamefont {Koepfli}, \citenamefont {Habegger}, \citenamefont
  {Eppenberger} \emph {et~al.}}]{horst2022transparent}%
  \BibitemOpen
  \bibfield  {author} {\bibinfo {author} {\bibfnamefont {Y.}~\bibnamefont
  {Horst}}, \bibinfo {author} {\bibfnamefont {T.}~\bibnamefont {Blatter}},
  \bibinfo {author} {\bibfnamefont {L.}~\bibnamefont {Kulmer}}, \bibinfo
  {author} {\bibfnamefont {B.~I.}\ \bibnamefont {Bitachon}}, \bibinfo {author}
  {\bibfnamefont {B.}~\bibnamefont {Baeuerle}}, \bibinfo {author}
  {\bibfnamefont {M.}~\bibnamefont {Destraz}}, \bibinfo {author} {\bibfnamefont
  {W.}~\bibnamefont {Heni}}, \bibinfo {author} {\bibfnamefont {S.}~\bibnamefont
  {Koepfli}}, \bibinfo {author} {\bibfnamefont {P.}~\bibnamefont {Habegger}},
  \bibinfo {author} {\bibfnamefont {M.}~\bibnamefont {Eppenberger}}, \emph
  {et~al.},\ }\bibfield  {title} {\bibinfo {title} {Transparent
  optical-thz-optical link at 240/192 gbit/s over 5/115 m enabled by
  plasmonics},\ }\href@noop {} {\bibfield  {journal} {\bibinfo  {journal}
  {Journal of Lightwave Technology}\ }\textbf {\bibinfo {volume} {40}},\
  \bibinfo {pages} {1690} (\bibinfo {year} {2022})}\BibitemShut {NoStop}%
\bibitem [{\citenamefont {Wijayanto}\ \emph {et~al.}(2013)\citenamefont
  {Wijayanto}, \citenamefont {Murata},\ and\ \citenamefont
  {Okamura}}]{wijayanto2013electrooptic}%
  \BibitemOpen
  \bibfield  {author} {\bibinfo {author} {\bibfnamefont {Y.~N.}\ \bibnamefont
  {Wijayanto}}, \bibinfo {author} {\bibfnamefont {H.}~\bibnamefont {Murata}},\
  and\ \bibinfo {author} {\bibfnamefont {Y.}~\bibnamefont {Okamura}},\
  }\bibfield  {title} {\bibinfo {title} {Electrooptic
  millimeter-wave--lightwave signal converters suspended to gap-embedded patch
  antennas on low-$ k $ dielectric materials},\ }\href@noop {} {\bibfield
  {journal} {\bibinfo  {journal} {IEEE Journal of Selected Topics in Quantum
  Electronics}\ }\textbf {\bibinfo {volume} {19}},\ \bibinfo {pages} {33}
  (\bibinfo {year} {2013})}\BibitemShut {NoStop}%
\bibitem [{\citenamefont {Kaji}\ \emph {et~al.}(2021)\citenamefont {Kaji},
  \citenamefont {Morohashi}, \citenamefont {Tominari}, \citenamefont {Sekine},
  \citenamefont {Yamada},\ and\ \citenamefont {Otomo}}]{kaji2021w}%
  \BibitemOpen
  \bibfield  {author} {\bibinfo {author} {\bibfnamefont {T.}~\bibnamefont
  {Kaji}}, \bibinfo {author} {\bibfnamefont {I.}~\bibnamefont {Morohashi}},
  \bibinfo {author} {\bibfnamefont {Y.}~\bibnamefont {Tominari}}, \bibinfo
  {author} {\bibfnamefont {N.}~\bibnamefont {Sekine}}, \bibinfo {author}
  {\bibfnamefont {T.}~\bibnamefont {Yamada}},\ and\ \bibinfo {author}
  {\bibfnamefont {A.}~\bibnamefont {Otomo}},\ }\bibfield  {title} {\bibinfo
  {title} {W-band optical modulators using electro-optic polymer waveguides and
  patch antenna arrays},\ }\href@noop {} {\bibfield  {journal} {\bibinfo
  {journal} {Optics Express}\ }\textbf {\bibinfo {volume} {29}},\ \bibinfo
  {pages} {29604} (\bibinfo {year} {2021})}\BibitemShut {NoStop}%
\bibitem [{\citenamefont {Murata}\ \emph {et~al.}(2021)\citenamefont {Murata},
  \citenamefont {Yokohashi}, \citenamefont {Matsukawa}, \citenamefont {Sato},
  \citenamefont {Onizawa},\ and\ \citenamefont {Kurokawa}}]{murata2021antenna}%
  \BibitemOpen
  \bibfield  {author} {\bibinfo {author} {\bibfnamefont {H.}~\bibnamefont
  {Murata}}, \bibinfo {author} {\bibfnamefont {H.}~\bibnamefont {Yokohashi}},
  \bibinfo {author} {\bibfnamefont {S.}~\bibnamefont {Matsukawa}}, \bibinfo
  {author} {\bibfnamefont {M.}~\bibnamefont {Sato}}, \bibinfo {author}
  {\bibfnamefont {M.}~\bibnamefont {Onizawa}},\ and\ \bibinfo {author}
  {\bibfnamefont {S.}~\bibnamefont {Kurokawa}},\ }\bibfield  {title} {\bibinfo
  {title} {Antenna-coupled electrode electro-optic modulator for 5g mobile
  applications},\ }\href@noop {} {\bibfield  {journal} {\bibinfo  {journal}
  {IEEE Journal of Microwaves}\ }\textbf {\bibinfo {volume} {1}},\ \bibinfo
  {pages} {902} (\bibinfo {year} {2021})}\BibitemShut {NoStop}%
\bibitem [{\citenamefont {Park}\ \emph {et~al.}(2015)\citenamefont {Park},
  \citenamefont {Pag{\'a}n}, \citenamefont {Murphy}, \citenamefont {Luo},
  \citenamefont {Jen},\ and\ \citenamefont {Herman}}]{park2015free}%
  \BibitemOpen
  \bibfield  {author} {\bibinfo {author} {\bibfnamefont {D.}~\bibnamefont
  {Park}}, \bibinfo {author} {\bibfnamefont {V.}~\bibnamefont {Pag{\'a}n}},
  \bibinfo {author} {\bibfnamefont {T.}~\bibnamefont {Murphy}}, \bibinfo
  {author} {\bibfnamefont {J.}~\bibnamefont {Luo}}, \bibinfo {author}
  {\bibfnamefont {A.-Y.}\ \bibnamefont {Jen}},\ and\ \bibinfo {author}
  {\bibfnamefont {W.}~\bibnamefont {Herman}},\ }\bibfield  {title} {\bibinfo
  {title} {Free space millimeter wave-coupled electro-optic high speed
  nonlinear polymer phase modulator with in-plane slotted patch antennas},\
  }\href@noop {} {\bibfield  {journal} {\bibinfo  {journal} {Optics express}\
  }\textbf {\bibinfo {volume} {23}},\ \bibinfo {pages} {9464} (\bibinfo {year}
  {2015})}\BibitemShut {NoStop}%
\bibitem [{\citenamefont {Miyazeki}\ \emph {et~al.}(2020)\citenamefont
  {Miyazeki}, \citenamefont {Yokohashi}, \citenamefont {Kodama}, \citenamefont
  {Murata},\ and\ \citenamefont {Arakawa}}]{miyazeki2020ingaas}%
  \BibitemOpen
  \bibfield  {author} {\bibinfo {author} {\bibfnamefont {Y.}~\bibnamefont
  {Miyazeki}}, \bibinfo {author} {\bibfnamefont {H.}~\bibnamefont {Yokohashi}},
  \bibinfo {author} {\bibfnamefont {S.}~\bibnamefont {Kodama}}, \bibinfo
  {author} {\bibfnamefont {H.}~\bibnamefont {Murata}},\ and\ \bibinfo {author}
  {\bibfnamefont {T.}~\bibnamefont {Arakawa}},\ }\bibfield  {title} {\bibinfo
  {title} {Ingaas/inalas multiple-quantum-well optical modulator integrated
  with a planar antenna for a millimeter-wave radio-over-fiber system},\
  }\href@noop {} {\bibfield  {journal} {\bibinfo  {journal} {Optics express}\
  }\textbf {\bibinfo {volume} {28}},\ \bibinfo {pages} {11583} (\bibinfo {year}
  {2020})}\BibitemShut {NoStop}%
\bibitem [{\citenamefont {Haykin}(1988)}]{haykin1988digital}%
  \BibitemOpen
  \bibfield  {author} {\bibinfo {author} {\bibfnamefont {S.}~\bibnamefont
  {Haykin}},\ }\bibfield  {title} {\bibinfo {title} {Digital communications},\
  }\href@noop {} {\bibfield  {journal} {\bibinfo  {journal} {New York}\ }
  (\bibinfo {year} {1988})}\BibitemShut {NoStop}%
\bibitem [{\citenamefont {Ghosh}\ and\ \citenamefont
  {Pendharker}(2021)}]{Ghosh_2021}%
  \BibitemOpen
  \bibfield  {author} {\bibinfo {author} {\bibfnamefont {N.}~\bibnamefont
  {Ghosh}}\ and\ \bibinfo {author} {\bibfnamefont {S.}~\bibnamefont
  {Pendharker}},\ }\bibfield  {title} {\bibinfo {title} {Wireless to optical
  phase mapping in a seamless digital wireless-photonic link},\ }\href@noop {}
  {\bibfield  {journal} {\bibinfo  {journal} {Journal of Optics}\ }\textbf
  {\bibinfo {volume} {23}},\ \bibinfo {pages} {125702} (\bibinfo {year}
  {2021})}\BibitemShut {NoStop}%
\bibitem [{\citenamefont {{N. Ghosh and S. Pendharker}}(2023)}]{10107736}%
  \BibitemOpen
  \bibfield  {author} {\bibinfo {author} {\bibnamefont {{N. Ghosh and S.
  Pendharker}}},\ }\bibfield  {title} {\bibinfo {title} {Electro-optic
  beamforming for wireless-to-optical seamless constellation-mapping},\ }\href
  {https://doi.org/10.1109/JLT.2023.3269812} {\bibfield  {journal} {\bibinfo
  {journal} {Journal of Lightwave Technology}\ }\textbf {\bibinfo {volume}
  {41}},\ \bibinfo {pages} {5851} (\bibinfo {year} {2023})}\BibitemShut
  {NoStop}%
\bibitem [{\citenamefont {Ghosh}\ and\ \citenamefont
  {Pendharker}(2024)}]{10535154}%
  \BibitemOpen
  \bibfield  {author} {\bibinfo {author} {\bibfnamefont {N.}~\bibnamefont
  {Ghosh}}\ and\ \bibinfo {author} {\bibfnamefont {S.}~\bibnamefont
  {Pendharker}},\ }\bibfield  {title} {\bibinfo {title} {Seamless digital
  wireless-photonic receiver with optical phase diversity},\ }\href
  {https://doi.org/10.1109/JLT.2024.3403231} {\bibfield  {journal} {\bibinfo
  {journal} {Journal of Lightwave Technology}\ ,\ \bibinfo {pages} {1}}
  (\bibinfo {year} {2024})}\BibitemShut {NoStop}%
\bibitem [{\citenamefont {Capmany}\ and\ \citenamefont
  {Fern{\'a}ndez-Pousa}(2010)}]{capmany2010quantum}%
  \BibitemOpen
  \bibfield  {author} {\bibinfo {author} {\bibfnamefont {J.}~\bibnamefont
  {Capmany}}\ and\ \bibinfo {author} {\bibfnamefont {C.~R.}\ \bibnamefont
  {Fern{\'a}ndez-Pousa}},\ }\bibfield  {title} {\bibinfo {title} {Quantum model
  for electro-optical phase modulation},\ }\href@noop {} {\bibfield  {journal}
  {\bibinfo  {journal} {JOSA B}\ }\textbf {\bibinfo {volume} {27}},\ \bibinfo
  {pages} {A119} (\bibinfo {year} {2010})}\BibitemShut {NoStop}%
\bibitem [{\citenamefont {Miroshnichenko}\ \emph {et~al.}(2017)\citenamefont
  {Miroshnichenko}, \citenamefont {Kiselev}, \citenamefont {Trifanov},\ and\
  \citenamefont {Gleim}}]{miroshnichenko2017algebraic}%
  \BibitemOpen
  \bibfield  {author} {\bibinfo {author} {\bibfnamefont {G.~P.}\ \bibnamefont
  {Miroshnichenko}}, \bibinfo {author} {\bibfnamefont {A.~D.}\ \bibnamefont
  {Kiselev}}, \bibinfo {author} {\bibfnamefont {A.~I.}\ \bibnamefont
  {Trifanov}},\ and\ \bibinfo {author} {\bibfnamefont {A.~V.}\ \bibnamefont
  {Gleim}},\ }\bibfield  {title} {\bibinfo {title} {Algebraic approach to
  electro-optic modulation of light: exactly solvable multimode quantum
  model},\ }\href@noop {} {\bibfield  {journal} {\bibinfo  {journal} {JOSA B}\
  }\textbf {\bibinfo {volume} {34}},\ \bibinfo {pages} {1177} (\bibinfo {year}
  {2017})}\BibitemShut {NoStop}%
\bibitem [{\citenamefont {Horoshko}\ \emph {et~al.}(2018)\citenamefont
  {Horoshko}, \citenamefont {Eskandary},\ and\ \citenamefont
  {Kilin}}]{horoshko2018quantum}%
  \BibitemOpen
  \bibfield  {author} {\bibinfo {author} {\bibfnamefont {D.}~\bibnamefont
  {Horoshko}}, \bibinfo {author} {\bibfnamefont {M.}~\bibnamefont
  {Eskandary}},\ and\ \bibinfo {author} {\bibfnamefont {S.~Y.}\ \bibnamefont
  {Kilin}},\ }\bibfield  {title} {\bibinfo {title} {Quantum model for
  traveling-wave electro-optical phase modulator},\ }\href@noop {} {\bibfield
  {journal} {\bibinfo  {journal} {JOSA B}\ }\textbf {\bibinfo {volume} {35}},\
  \bibinfo {pages} {2744} (\bibinfo {year} {2018})}\BibitemShut {NoStop}%
\bibitem [{\citenamefont {Pratap}\ and\ \citenamefont
  {Ramachandran}(2020)}]{pratap2020quantum}%
  \BibitemOpen
  \bibfield  {author} {\bibinfo {author} {\bibfnamefont {R.}~\bibnamefont
  {Pratap}}\ and\ \bibinfo {author} {\bibfnamefont {H.}~\bibnamefont
  {Ramachandran}},\ }\bibfield  {title} {\bibinfo {title} {Quantum analysis of
  a traveling-wave electro-optic phase modulator in the presence of the phase
  noise from a radio frequency oscillator and a laser},\ }\href@noop {}
  {\bibfield  {journal} {\bibinfo  {journal} {JOSA B}\ }\textbf {\bibinfo
  {volume} {37}},\ \bibinfo {pages} {3016} (\bibinfo {year}
  {2020})}\BibitemShut {NoStop}%
\bibitem [{\citenamefont {Pratap}\ and\ \citenamefont
  {Ramachandran}(2023)}]{pratap2023photon}%
  \BibitemOpen
  \bibfield  {author} {\bibinfo {author} {\bibfnamefont {R.}~\bibnamefont
  {Pratap}}\ and\ \bibinfo {author} {\bibfnamefont {H.}~\bibnamefont
  {Ramachandran}},\ }\bibfield  {title} {\bibinfo {title} {Photon statistics
  and quantum phase distribution for coherent optical link inthe presence of
  phasenoise from both laser source and rf oscillator, and dispersion},\
  }\href@noop {} {\bibfield  {journal} {\bibinfo  {journal} {Journal of
  Optics}\ } (\bibinfo {year} {2023})}\BibitemShut {NoStop}%
\bibitem [{\citenamefont {Imany}\ \emph {et~al.}(2018)\citenamefont {Imany},
  \citenamefont {Odele}, \citenamefont {Jaramillo-Villegas}, \citenamefont
  {Leaird},\ and\ \citenamefont {Weiner}}]{imany2018characterization}%
  \BibitemOpen
  \bibfield  {author} {\bibinfo {author} {\bibfnamefont {P.}~\bibnamefont
  {Imany}}, \bibinfo {author} {\bibfnamefont {O.~D.}\ \bibnamefont {Odele}},
  \bibinfo {author} {\bibfnamefont {J.~A.}\ \bibnamefont {Jaramillo-Villegas}},
  \bibinfo {author} {\bibfnamefont {D.~E.}\ \bibnamefont {Leaird}},\ and\
  \bibinfo {author} {\bibfnamefont {A.~M.}\ \bibnamefont {Weiner}},\ }\bibfield
   {title} {\bibinfo {title} {Characterization of coherent quantum frequency
  combs using electro-optic phase modulation},\ }\href@noop {} {\bibfield
  {journal} {\bibinfo  {journal} {Physical Review A}\ }\textbf {\bibinfo
  {volume} {97}},\ \bibinfo {pages} {013813} (\bibinfo {year}
  {2018})}\BibitemShut {NoStop}%
\bibitem [{\citenamefont {Gerry}\ and\ \citenamefont
  {Knight}(2023)}]{gerry2023introductory}%
  \BibitemOpen
  \bibfield  {author} {\bibinfo {author} {\bibfnamefont {C.~C.}\ \bibnamefont
  {Gerry}}\ and\ \bibinfo {author} {\bibfnamefont {P.~L.}\ \bibnamefont
  {Knight}},\ }\href@noop {} {\emph {\bibinfo {title} {Introductory quantum
  optics}}}\ (\bibinfo  {publisher} {Cambridge university press},\ \bibinfo
  {year} {2023})\BibitemShut {NoStop}%
\bibitem [{\citenamefont {Bransden}(2000)}]{bransden2000quantum}%
  \BibitemOpen
  \bibfield  {author} {\bibinfo {author} {\bibfnamefont {B.~H.}\ \bibnamefont
  {Bransden}},\ }\href@noop {} {\emph {\bibinfo {title} {Quantum mechanics}}}\
  (\bibinfo  {publisher} {Pearson Education India},\ \bibinfo {year}
  {2000})\BibitemShut {NoStop}%
\bibitem [{\citenamefont {Furusawa}\ and\ \citenamefont
  {Furusawa}(2015)}]{furusawa2015quantum}%
  \BibitemOpen
  \bibfield  {author} {\bibinfo {author} {\bibfnamefont {A.}~\bibnamefont
  {Furusawa}}\ and\ \bibinfo {author} {\bibfnamefont {A.}~\bibnamefont
  {Furusawa}},\ }\href@noop {} {\emph {\bibinfo {title} {Quantum states of
  light}}}\ (\bibinfo  {publisher} {Springer},\ \bibinfo {year}
  {2015})\BibitemShut {NoStop}%
\bibitem [{\citenamefont {Asavanant}\ and\ \citenamefont
  {Furusawa}(2022)}]{asavanant2022optical}%
  \BibitemOpen
  \bibfield  {author} {\bibinfo {author} {\bibfnamefont {W.}~\bibnamefont
  {Asavanant}}\ and\ \bibinfo {author} {\bibfnamefont {A.}~\bibnamefont
  {Furusawa}},\ }\href@noop {} {\emph {\bibinfo {title} {Optical Quantum
  Computers}}}\ (\bibinfo  {publisher} {AIP Publishing LLC},\ \bibinfo {year}
  {2022})\BibitemShut {NoStop}%
\end{thebibliography}
%\bibliographystyle{style=numeric-comp}

%\providecommand{\noopsort}[1]{}\providecommand{\singleletter}[1]{#1}%

\providecommand{\noopsort}[1]{}\providecommand{\singleletter}[1]{#1}%

\end{document}